\documentclass[aps,twocolumn,%superscriptaddress,showpacs,showkeys,
groupedaddress,nofootinbib,notitlepage,nobalancelastpage,nobibnotes,preprintnumbers]{revtex4}
\pdfoutput=1
\usepackage{amsmath,amsfonts,amssymb,mathrsfs,graphicx}
\usepackage[squaren]{SIunits}
\usepackage{verbatim}
\usepackage[utf8]{inputenc}
\usepackage{float}
\usepackage{longtable}
\usepackage{bbold}
\usepackage{pifont}
\usepackage{hyperref}
\usepackage{url}
\usepackage[dvipsnames]{xcolor}

\newcommand{\ie}{{\it i.e.}}

\newcommand{\eg}{{\it e.g.}}

\newcommand{\cf}{{\it cf.}}

\newcommand{\eq}{Eq.}

\newcommand{\fig}{Fig.}
\newcommand{\Fig}{Fig.}

\newcommand{\Ref}{Ref.}
\newcommand{\Refs}{Refs.}

\newcommand{\Tab}{Tab.}

\newcommand{\ba}{\begin{array}}
\newcommand{\ea}{\end{array}}

\newcommand{\equ}[1]{\eq~(\ref{equ:#1})}
\newcommand{\figu}[1]{\fig~\ref{fig:#1}}

\newcommand{\bi}{\begin{itemize}}
\newcommand{\ei}{\end{itemize}}

\begin{document}

\preprint{DESY 16-136} 
\vspace*{0.7cm}
\title{Perspectives for tests of neutrino mass generation at the GeV scale: \newline Experimental reach versus theoretical predictions}

\author{Rasmus W. Rasmussen}
\email{rasmus.westphal.rasmussen@desy.de}
\affiliation{DESY, Platanenallee 6, 15738 Zeuthen, Germany}

\author{Walter Winter}
\email{walter.winter@desy.de}
\affiliation{DESY, Platanenallee 6, 15738 Zeuthen, Germany}

\date{\today}

\begin{abstract}
We discuss the parameter space reach of future experiments searching for heavy neutral leptons (HNLs). We consider the GeV seesaw model with three HNL generations and focus on two classes of models: generic assumptions (such as random mass matrices or the Casas-Ibarra parametrization) and flavor symmetry-generated models. We demonstrate that the generic approaches lead to comparable parameter space predictions, which tend to be at least partially within the reach of future experiments. On the other hand, specific flavor symmetry models yield more refined predictions, and some of these can be more clearly excluded. We also highlight the importance of measuring the flavor-dependent couplings of the HNLs as a model discriminator, and we clarify the impact of assumptions frequently used in the literature to show the parameter space reach for the active-sterile mixings. 
\end{abstract}

\maketitle

\section{Introduction}

The discovery of massive neutrinos through neutrino oscillations implies evidence for physics beyond the Standard Model (SM). Theoretical challenges include (i) an understanding of the smallness of neutrino masses and (ii) describing the vastly different pattern of mixing among neutrinos from the quarks. For example, a key question is whether it is possible to reconcile the large neutrino mixing with small quark mixing in theories beyond the SM such as theories of Grand Unification. A natural explanation for the smallness of neutrino masses is the so-called seesaw mechanism~\cite{MINKOWSKI1977421, GellMann:1980vs, Yanagida01091980, PhysRevLett.44.912, PhysRevD.22.2227}, where the interaction with a heavy partner  suppresses the neutrino mass scale. The energy scale associated with the heavy partner can range from the sub-eV scale to the Grand Unified Theory (GUT) scale~\cite{Abazajian:2012ys}. The main scenarios studied in the literature associate right-handed neutrinos  with the eV, GeV, TeV, or GUT scale. For example, the reactor and the LSND anomalies in the neutrino oscillation data~\cite{Athanassopoulos:1996ds, Mention:2011rk} point toward the existence of one (or multiple) sterile neutrinos with a mass around 1 eV~\cite{Babu:2016fdt, Giunti:2015wnd}. If the right-handed neutrinos have masses at the GeV scale, the baryon asymmetry can be described together with neutrino masses and mixings. TeV-scale sterile neutrinos can be found by current or future collider experiments~\cite{Asaka:2015oia, Banerjee:2015gca, Antusch:2016vyf, ATLAS:2012ak, Khachatryan:2014dka, Das:2012ze, Das:2014jxa}. Grand Unified Theories usually predict right-handed neutrinos which makes them able to explain the baryon asymmetry via leptogenesis~\cite{Ishihara:2015uua, Bando:2004hi}. Additionally, HNLs have also been studied as a dark matter candidate with a mass in the keV or PeV range~\cite{Campos:2016gjh, Dev:2016qbd, Daikoku:2015vsa, Adhikari:2016bei, Merle:2013gea, Heurtier:2016iac}. For a full review concerning right-handed neutrinos at energy scales mention above and their consequence, see Refs.~\cite{Drewes:2013gca, Alekhin:2015byh}. 

The possible mixing between the left-handed and right-handed neutrinos can be studied experimentally. Accelerator-based experiments probe the mass range MeV-TeV depending on the center-of-mass energy available and the process investigated~\cite{Vaitaitis:1999wq, Artamonov:2014urb, Vilain1995387, Das:2015toa, Das:2016akd}, just to mention some examples. Additionally, global analyses have investigated the parameter space of the active-sterile mixing in the GeV-TeV range by the use of electroweak (EW) precision observables~\cite{Mohapatra2016423, deGouvea:2015euy, Akhmedov:2013hec, Ibarra:2011xn, Ibarra:2010xw, Fernandez-Martinez:2016lgt}.

On the theory side, several studies in the literature have investigated the parameter space of the seesaw mechanism. Besides explaining small neutrino masses, other phenomena can be realized with the right-handed neutrinos, such as baryon asymmetry and dark matter~\cite{Akhmedov:1998qx, Canetti:2012kh,  Escudero:2016tzx, DiBari:2016guw, Parida:2016asc}. An interesting scenario is the so-called neutrino Minimal Standard Model ($\nu$MSM)~\cite{Asaka:2005an, Asaka:2005pn}, for which the parameter space of the active-sterile mixing is constrained by the following requirements. One sterile neutrino is a dark matter candidate with mass ${M_{I} \sim}$~keV and total active-sterile mixing ${|U_{I}|^{2} \lesssim 10^{-8}}$. Two other sterile neutrinos are degenerate with a mass at the GeV scale to describe both neutrino masses and baryon asymmetry. They can be searched for by future experiments. Recently, different authors have investigated neutrino mixings, neutrinoless double beta decay and baryon asymmetry in the context of future experiments being able to probe the $\nu$MSM or the scenario with three sterile neutrinos at the GeV scale~\cite{Canetti:2014dka, Drewes:2015iva, Hambye:2016sby, Drewes:2016lqo, Asaka:2016zib, Drewes:2016gmt, Hernandez:2016kel}. Note that three sterile neutrinos at the GeV scale (compared to two neutrinos in the $\nu$MSM) imply that no fine-tuning of the sterile neutrino mass is needed to describe the baryon asymmetry~\cite{Drewes:2012ma} and that a much larger parameter space is allowed. A dark matter candidate can be easily added as another generation without consequence if it is weakly mixing. An alternative model to the $\nu$MSM is presented in \Ref~\cite{DiBari:2016guw} which also explains neutrino mixing, dark matter and baryon asymmetry by introducing nonstandard interactions among the sterile neutrinos in the model.

In order to obtain predictions for the neutrino mass and mixings from fundamental principles, flavor symmetries have been proposed; see \eg\ \Ref~\cite{King:2013eh} for a review. Various models have been shown to describe the neutrino oscillation parameters within their $3\sigma$ range~\cite{King:2016pgv, Kanemura:2016ixx, Fonseca:2014lfa, Neder:2015wsa, Yao:2015dwa, Parattu:2010cy} using a variety of flavor symmetries such as $A_{5}$~\cite{Li:2015jxa, Ballett:2015wia, DiIura:2015kfa, Turner:2015uta, Joshipura:2016hvn}, $S_{4}$~\cite{Bazzocchi:2012st, Krishnan:2012me, Luhn:2013vna, Vien:2016jkz, King:2016yvg}, $A_{4}$~\cite{1742-6596-627-1-012003, Pramanick:2015qga, Morisi:2013qna, Felipe:2013vwa, Amitai:2012em, Hernandez:2015tna, King:2011zj}, and $\Delta$-groups such as $\Delta(27)$, $\Delta(48)$, $\Delta(54)$, $\Delta(6n^2)$ and $\Delta(3n^2)$ for $n \in \mathbb{N}$~\cite{King:2014rwa, Vien:2016tmh, Ding:2014ora, Hagedorn:2014wha, Ding:2014hva, Hernandez:2016eod, Abbas:2015zna, Varzielas:2015aua, Chulia:2016giq, Carballo-Perez:2016ooy}. Additional phenomena, such as leptogenesis, can be also be explained~\cite{Lucente:2016zbs, Gu:2015wkd, Cline:2015yba, Karmakar:2015jza, Abada:2015rta} in such models. From a different perspective,  the structure of the mass matrices and their consequences without considering the origin in terms of a  flavor symmetry have been considered; see \eg\ \Refs~\cite{doi:10.1142/S0217732315501382, Lavoura:2015wwa, Singh:2016qcf, Kitabayashi:2016fsz, Borah:2016xkc}. Many models give a prediction of the unknown observables, such as CP-violating phases, the absolute neutrino mass scale, neutrino mass ordering, the nature of neutrinos (Dirac or Majorana particles), and active-sterile mixing. This means that they can be distinguished and possibly excluded by future experiments measuring the observables. We will use flavor symmetries to predict the structure of the heavy neutrino mass matrix and the Yukawa couplings~\cite{Plentinger:2008up}, which will have consequences for the active-sterile mixings.
 
In this work, we study if it is possible to distinguish among different theoretical predictions for the active-sterile mixing when introducing $\mathcal{N}=3$ sterile neutrinos in a low-scale (GeV) seesaw model (in comparison to the usual choice of a GUT seesaw model). We study the total and flavor-dependent mixings of the sterile neutrinos, both in the model-independent and the flavor symmetry contexts. We also discuss the corresponding exclusion bounds of the parameter space. We consider normal ordering only, since the flavor models in this study were produced under this assumption.

\section{Heavy neutral leptons in theory and experiment}

In this section we sketch the seesaw mechanism with HNLs including the physical observables, we discuss the  experimental signatures, and we introduce the planned future experiments which are the main focus of this study.

\subsection{Theoretical framework}

The basic idea of the seesaw (type~I) mechanism is to minimally extend the SM by heavy right-handed neutrinos in order to describe the smallness of neutrino mass. Integrating out the heavy fields leads to light Majorana neutrino masses after electroweak symmetry breaking (EWSB). The mass term Lagrangian in the seesaw mechanism is, below the EWSB scale, given by~\cite{Drewes:2013gca}
\begin{align}
 -\mathcal{L}_{\text{mass}} & =\frac{1}{2}N^{c}_{R}M_{R}N_{R}+\nu_{L}M_{D}N_{R}+h.c. \nonumber \\
& =\frac{1}{2}
\begin{pmatrix}
 \nu_{L} & N^{c}_{R}
\end{pmatrix}
\begin{pmatrix}
 0 & M_{D} \\
M_{D}^{T} & M_{R}
\end{pmatrix}
\begin{pmatrix}
 \nu^{c}_{L} \\
N_{R}
\end{pmatrix} + h.c.
\label{equ:massseesaw}
\end{align}
where ${\nu_{L} \ (N_{R})}$ is the left-handed (right-handed) neutrino and ${M_{D} \ (M_{R})}$ is the Dirac (Majorana) mass matrix with the overall mass scale ${m_{D} \ (m_{R})}$. By block diagonalizing the mass matrix in \equ{massseesaw}, the effective mass matrices for the light and heavy neutrinos are obtained. This introduces the mixing angle ${\theta=M_{D}M_{R}^{-1}}$ which has to satisfy $\theta \ll 1$ such that the diagonalizing matrix is unitary [up to ${\mathcal{O}(\theta^{4})}$]~\cite{Drewes:2013gca}. The effective neutrino mass matrices of the light and heavy neutrinos are 
\begin{equation}
 m_{\nu}=-M_{D}M_{R}^{-1}M_{D}^{T} \hspace*{0.5cm} \text{and}  \hspace*{0.5cm} M_{N}=M_{R} \, ,
\label{equ:seesaw}
\end{equation}
neglecting higher order terms. This is the famous seesaw formula where the suppression from the Majorana mass matrix describes the smallness of the neutrino masses. The mass matrices can be diagonalized by the unitary matrices $U_{\nu}$ and $U_{N}$ according to
\begin{equation}
 U_{\nu}^{\dagger}m_{\nu}U_{\nu}^{*}=\text{diag}(m_{1}, m_{2}, m_{3}) 
\end{equation}
and
\begin{equation}
 U_{N}^{\dagger}M_{N}U_{N}^{*}=\text{diag}(M_{1}, M_{2}, M_{3})
\end{equation}
where ${m_{i} \ (M_{I})}$ are the masses of the light (heavy) neutrinos. The mixing matrix describing the mixing in the charged current is then given by ${V \equiv U_{\ell}^{\dagger}\left(1-\frac{1}{2}\theta\theta^{\dagger}\right)U_{\nu}}$, where $U_{\ell}$ comes from diagonalizing the charged lepton Yukawa couplings $Y_\ell$ (or mass matrix). For our purposes,  the term ${\frac{1}{2}\theta\theta^{\dagger}}$ is negligible as ${\theta \ll 1}$, implying that the mixing  matrix becomes the unitary PMNS matrix, \ie,  ${V = U_{\text{PMNS}} \equiv U_{\ell}^{\dagger}U_{\nu}}$. 
We adopt the standard parametrization for $U_{\text{PMNS}}$~\cite{Agashe:2014kda}. Other observables in the seesaw context describe the active-sterile mixing~\cite{Drewes:2013gca, Drewes:2015iva}
\begin{equation}
 |U_{\alpha I}|^{2}=|(U_{\ell}^{\dagger}\theta U_{N})_{\alpha I}|^{2} \hspace*{0.3cm} \text{and} \hspace*{0.3cm}|U_{I}|^{2}=\sum_{\alpha}|U_{\alpha I}|^{2}
\label{equ:activesterilemixing}
\end{equation}
where $|U_{\alpha I}|^{2}$ represents the individual mixing element of a sterile neutrino ${I=\{1,2,3\}}$ with a particular flavor ${\alpha=\{e,\mu,\tau\}}$, whereas $|U_{I}|^{2}$ is the total mixing for the sterile neutrino $I$.

\subsection{Experimental signatures}

 The active-sterile mixings enter physical observables such as the decay rates~\cite{Gorbunov:2007ak, Canetti:2012kh} 
\begin{align}
\begin{aligned}
 &\Gamma(N_{I} \rightarrow X\ell_{\alpha} )  = \frac{|U_{\alpha I}|^{2}}{16 \pi}G_{f}^{2}|V_{X}|^{2}f_{X}^{2}M_{I}^{3} \\ & \times \left(\left(1-\frac{M_{\ell}^{2}}{M_{I}^{2}} \right)^{2}-\frac{M_{X}^{2}}{M_{I}^{2}}\left(1+\frac{M_{\ell}^{2}}{M_{I}^{2}} \right) \right) \\ &  \times \sqrt{\left(1-\frac{(M_{X}-M_{\ell})^{2}}{M_{I}^{2}}\right)\left(1-\frac{(M_{X}+M_{\ell})^{2}}{M_{I}^{2}}\right)} \, ,
\label{equ:chargedhadrondecay}
\end{aligned}
\end{align}
where $X$ is a charged hadron with mass $M_{X}$, $G_{f}$ is the Fermi coupling constant and ${M_{I} \ (M_{\ell})}$ is the mass of the sterile neutrino (charged lepton). The CKM matrix element $V_{X}$ and the decay constant $f_{X}$ of the charged hadron are also present in the equation. A pion as the charged hadron means ${|V_{X}|^{2}=|V_{ud}|^{2} \equiv 0.949}$ and ${f_{X}=f_{\pi}\equiv 130.0}$~MeV, whereas ${|V_{X}|^{2}=|V_{us}|^{2} \equiv 0.051}$ and ${f_{X}=f_{K}\equiv 159.8}$~MeV for a kaon in the final state~\cite{Agashe:2014kda}, just to mention some examples. There are other experimental signatures of sterile neutrinos besides those already mentioned. The active-sterile mixing can modify the EW precision observables such as the Z invisible decay width, the Fermi constant $G_{F}$, and other EW parameters in the SM and lead to lepton number/flavor violation (LFV)~\cite{Antusch:2014woa, Antusch:2006vwa, Abada:2007ux, Nardi:1994iv, Nardi:1994nw, Bergmann:1998rg, Abada:2012mc, Abada:2013aba, Asaka:2014kia}. Therefore, deviations from the SM values of these observables might suggest the existence of sterile neutrinos. The sterile neutrinos also contribute to the neutrinoless double beta decay amplitude; the parameter space of the effective mass $m_{\beta \beta}$ is larger than in the standard $3\nu$ case. This can be used  to probe the existence of a sterile neutrino, but it does not guarantee an observable $m_{\beta \beta}$ due to possible cancellations~\cite{Pascoli:2013fiz, Helo:2015fba}.\footnote{In the future, better limits from $m_{\beta \beta}$ and LFV may potentially constrain the active-sterile mixing even further. However, note that these are only indirect constraints. Take the double beta mass as an example, which can be written as~\cite{Drewes:2015iva}
\begin{equation}
 m_{\beta \beta}=\left| \sum_{i} (U_{\text{PMNS}})_{e i}^{2} m_{i} + \sum_{I} U_{e I}^{2} M_{A}^{2} F_{A} / M_{I} \right|
\end{equation}
where the first sum is the usual part from the $3\nu$ paradigm and the second sum involves the sterile neutrinos. Due to the large dimensionality (the extra freedom from each sterile neutrino), it is difficult to obtain direct constraints on the active-sterile mixings. We still apply the limits from $m_{\beta \beta}$ and LFV; however, we will focus on direct limits when comparing to our results.} Besides the decay in \equ{chargedhadrondecay}, the sterile neutrino can also decay into heavier particles such as charmed mesons, b-mesons, gauge bosons and Higgs bosons if kinematically allowed \cite{RichardJacobsson}. If the sterile neutrino is lighter, peak searches in meson decays can be preformed \cite{Deppisch:2015qwa}. Different types of experiments can study the production and decay of the sterile neutrinos. Beam-dump experiments are able to probe the mass range ${0.1-2}$~GeV, whereas B-factories can probe heavier sterile neutrino masses ${2-5}$~GeV. Beyond this range, hadron or lepton colliders can search for sterile neutrinos by investigating displaced vertices involving gauge bosons and Higgs bosons in the range ${5 \ \text{GeV}-3 \ \text{TeV}}$~\cite{RichardJacobsson, Helo:2013esa}. Additionally, all the experiments can investigate lepton flavor/number violating processes, such as ${K^{+} \rightarrow \ell^{+}\ell^{+}N}$ with ${\ell=\{ e, \mu\}}$~\cite{Deppisch:2015qwa}. Current experiments searching for sterile neutrinos across a wide range of energies ${(\text{MeV}-\text{TeV})}$ are, \eg\, BABAR, Belle, LHCb, ATLAS, and CMS~\cite{RichardJacobsson}. Some of the proposed experiments capable of searching for sterile neutrinos in the future in same energy range are Search for Hidden Particles (SHiP), NA62, ILC, CepC and Future Circular Collider (FCC)~\cite{RichardJacobsson}. 

\subsection{Future experiments}

In this work, we consider the SHiP~\cite{Anelli:2015pba, Alekhin:2015byh}, FCC~\cite{Blondel:2014bra}, and Deep Underground Neutrino Experiment (DUNE, formerly LBNE)~\cite{Adams:2013qkq, Acciarri:2015uup} experiments, which are sensitive to the active-sterile mixing in the GeV range. These are representative cases for the different proposed experiments which might be built in the future. SHiP is a proposed beam-dump experiment which is supposed to be situated at the SPS accelerator at CERN. This experiment is dedicated to search for hidden particles such as sterile neutrinos, dark photons, and supersymmetric particles. The final state decays investigated by the SHiP Collaboration are listed in Table~5.3 in \Ref~\cite{Anelli:2015pba}. From these decays, we can deduce the observables: the two-body decay ${N_{I} \rightarrow \mu \pi}$ is sensitive to $|U_{\mu I}|^{2}$ since the flavor of the charged lepton implies that the sterile neutrino must have mixed into a muon neutrino leading to the considered final decay (a similar argument holds for the decay $N_{I} \rightarrow e \pi$, which is sensitive to $|U_{e I}|^{2}$). The decay ${N_{I} \rightarrow \eta \nu \rightarrow \pi^{+} \pi^{-} \pi^{0} + p^{\text{miss}}}$ is sensitive to the total mixing $|U_{I}|^{2}$ because SHiP considers the light neutrino as missing energy, meaning that the flavor is not measured. Therefore the SHiP experiment can, in principle, measure the individual mixing elements $|U_{e I}|^{2}$ and $|U_{\mu I}|^{2}$, and the total mixing $|U_{I}|^{2}$. Since we are, however, not aware of any sensitivity study for the total mixing yet without assuming a ratio among the individual mixing elements, we will not display any direct bounds for the total mixing. Note that a bound for the total mixing could be derived from the bounds of the individual mixings if the decay with tau leptons in the final state were measured.

DUNE is a proposed long-baseline neutrino experiment at Fermilab with a baseline of 1300 km, where the far detector is located at the Sanford Research Facility Lab in South Dakota. Its primary goal is to measure the neutrino mass ordering and  the  leptonic CP-violating phase $\delta$. With the near detector, it will have sensitivity to the active-sterile mixing. The DUNE Collaboration has not yet reported the search modes it is investigating, but the experiment is sensitive to a mass range similar to that of SHiP. As the proton energy is lower than for SHiP\footnote{It is expected to be 80-120~GeV compared to 400 GeV at the SHiP experiment \cite{Adams:2013qkq, Acciarri:2015uup}.}, we do not expect significant sensitivity in searches involving tau leptons, and we assume that the same final states as for SHiP will be studied, specifically $|U_{e I}|^{2}$ and $|U_{\mu I}|^{2}$ (and $|U_{I}|^{2}$, for which we do not show any direct sensitivity curve). 

The FCC experiment is a proposed successor of the LHC experiments with an accelerator circumference of 80--100 km. As a first step in the physics program of the FCC, colliding leptons with a center-of-mass (CM) energy of 90--350 GeV are considered~\cite{Blondel:2014bra}, before colliding hadrons with CM energies up to 100 TeV. The FCC's main goal is study the Higgs boson's couplings at the percent level. However, it can also investigate the parameter space of active-sterile mixing in the GeV range. Specifically, one can measure the Z-boson partial decay width 
\begin{equation}
 \Gamma_{Z \rightarrow \nu N_{I}}=3\Gamma_{Z \rightarrow \nu \nu}^{\text{SM}}|U_{I}|^{2}(1-(M_{I}/M_{Z})^{2})^{2}(1+(M_{I}/M_{Z})^{2})
\label{equ:zboson}
\end{equation}
where ${\Gamma_{Z \rightarrow \nu \nu}^{\text{SM}}}$ is the SM decay rate of a Z-boson into two light neutrinos and $M_{I}$ is the mass of the sterile neutrino. It can be seen from \equ{zboson} that the FCC experiment is only sensitive to the total mixing $|U_{I}|^{2}$ at the Z-pole.\footnote{FCC and other proposed lepton colliders (ILC and CepC) can also measure the individual mixing element $|U_{eI}|^{2}$ if the center-of-mass energy is increased to $200-500$~GeV. However, these measurements are less sensitive than FCC on the total mixing. We therefore  disregard them since the individual mixing elements cannot violate the bound on the total mixing \cite{Antusch:2016vyf, RichardJacobsson}.} There are different experimental channels to be investigated at lepton colliders; here, we mention the two most promising ones. The channel ${e^{-}e^{+} \rightarrow N(\rightarrow \ell^{\mp}W^{\pm})\nu_{\ell}}$ leads to one lepton, two jets, and missing energy, where the hadronic activity from the jets can be controlled by kinematical cuts. Another channel ${e^{-}e^{+} \rightarrow N(\rightarrow \ell^{\prime \mp}W^{\pm})\ell^{\mp}W^{\pm}}$ leads to two leptons and four jets. Again, kinematical cuts and selecting two leptons with the same electric charge can reduce the background. Since the two outgoing leptons have the same sign, the process is lepton number violating and sensitive to the Majorana nature of the neutrinos \cite{Das:2016hof, RichardJacobsson}.

The sensitivity to the active-sterile mixing of the experiments has been studied under different assumptions for the ratios among the flavor-dependent mixings which makes it possible for the SHiP and DUNE collaborations to translate a bound on the individual mixing element into a bound on the total mixing. The FCC Collaboration only considers the total mixing; therefore no assumption of that kind is needed in their study. The SHiP collaboration obtained their sensitivity to the {\em total} active-sterile mixing for five different scenarios with these assumptions~\cite{Anelli:2015pba}:
\begin{align} 
&\text{\bf{Case 1:}} |U_{eI}|^{2}:|U_{\mu I}|^{2}:|U_{\tau I}|^{2} \sim 52 : 1 : 1, \text{IO} \nonumber \\ 
&\text{\bf{Case 2:}} |U_{eI}|^{2}:|U_{\mu I}|^{2}:|U_{\tau I}|^{2} \sim 1 : 16 : 3.8,  \text{NO} \nonumber \\ 
&\text{\bf{Case 3:}} |U_{eI}|^{2}:|U_{\mu I}|^{2}:|U_{\tau I}|^{2} \sim 0.061 : 1 : 4.3,  \text{NO} \label{equ:assumption}  \\ 
&\text{\bf{Case 4:}} |U_{eI}|^{2}:|U_{\mu I}|^{2}:|U_{\tau I}|^{2} \sim 48 : 1 : 1,  \text{IO} \nonumber  \\ 
&\text{\bf{Case 5:}} |U_{eI}|^{2}:|U_{\mu I}|^{2}:|U_{\tau I}|^{2} \sim 1 : 11 : 11,  \text{NO} \nonumber
\end{align}
where NO (IO) means normal (inverted) ordering of the light neutrinos. The first three scenarios imply that the sterile neutrinos predominantly mix with one flavor (electron, muon or tau)~\cite{Gorbunov:2007ak}, whereas the last two scenarios are interesting to generate a sufficient amount of baryon asymmetry~\cite{Canetti:2010aw}. Case~2 is SHiP's benchmark scenario, which means that the sensitivity to the total active-sterile mixing is calculated for this case only, whereas the sensitivity for the other cases has been obtained by extrapolating the sensitivity from case~2 by using the ratio among the individual mixing elements. The conclusions are derived from the decay ${N_{I} \rightarrow \mu\pi}$. Note, however, that  the SHiP Collaboration has also investigated  the processes ${N_{I} \rightarrow \mu\mu\nu}$ and ${N_{I} \rightarrow ee\nu}$ individually.

The DUNE Collaboration estimated their sensitivity curve~\cite{Adams:2013qkq} for the total mixing by scaling experimental parameters, such as protons on target, number of produced charm mesons, detector length, and detector area with the CHARM~\cite{Dorenbosch1986473} and the PS191~\cite{BERNARDI1988332} experiments. This means that the sensitivity curve is extrapolated to the DUNE experiment. However, CHARM and PS191 have only reported sensitivity bounds of the individual mixing elements $|U_{e I}|^{2}$ and $|U_{\mu I}|^{2}$, which means that DUNE's sensitivity curve is only valid when these individual mixing elements dominate  the total mixing. Therefore, the underlying assumption in terms of flavor observables are similar to the SHiP experiment. 

Several sensitivity bounds have been reported by each of the future experiments for changing the experiment parameters, such as detector length, running time, and decay length.  We use the more optimistic bounds presented by the experimental collaborations, \ie, the ones for a detector length of 30 m~\cite{Adams:2013qkq} and $10^{13}$ Z-bosons/decay length 0.01-500 mm~\cite{Blondel:2014bra} for the DUNE and FCC experiments, respectively. We only  consider the normal ordering, which means that, consequently, we use the active-sterile mixing in case~2 [\equ{assumption}] for the SHiP experiment, whenever applicable.

\section{Model-independent view of the parameter space}

Here two procedures are discussed which produce viable candidates for $Y_{\ell}, M_{D}$ and $M_{R}$ leading to neutrino physics observables in the allowed parameter ranges. The first procedure relies on the Casas-Ibarra parametrization starting from the observables as an input and parametrizing the degrees of freedom. In the second procedure, we vary the mass matrix entries randomly to generate neutrino oscillation parameters within their $3\sigma$ range. Then, we discuss the result obtained from these procedures and compare them to the sensitivity bound from the future experiments.

\subsection{Method}

The Dirac mass matrix can be constructed using  the Casas-Ibarra parametrization~\cite{Casas:2001sr}
\begin{equation}
 M_{D}=U_{\text{PMNS}}\sqrt{m_{\nu}^{\text{diag}}}\mathcal{R}^{T}\sqrt{M_{R}}
\label{equ:casibapara}
\end{equation}
separating the physical observables $U_{\text{PMNS}}$ and  the neutrino masses ${m_{\nu}^{\text{diag}}=\text{diag}(m_{1},m_{2},m_{3})}$ from the degrees of freedom not directly accessible ${M_{R}=\text{diag}(M_{1},M_{2},M_{3})}$ and $\mathcal{R}$ (see below). Here it is assumed that the charged lepton mass and heavy neutrino mass matrices are diagonal, which means that ${U_{\text{PMNS}} \equiv U_{\nu}}$ directly diagonalizes the light neutrino mass matrix. Note that these assumptions do not impose any restrictions with respect to the physical observables, but an underlying flavor symmetry may not be visible in that basis anymore. On the other hand, using \equ{casibapara} for the generation of $M_D$, the neutrino masses and mixings automatically match the predictions (as they are used as an input).

In order to generate possible models with GeV HNL masses, the neutrino oscillation parameters ($\theta_{ij}$ and $\Delta m_{ij}^{2}$) are chosen randomly from their $3\sigma$ ranges~\cite{GonzalezGarcia:2014bfa}. The Dirac and Majorana phases in the PMNS mixing matrix are chosen from the interval ${\delta, \alpha_{i} \in [0, 2\pi]}$. The lightest neutrino mass is chosen within the interval ${m_{\text{min}} \in [0,0.23] \ \text{eV}}$ to satisfy  the upper bound on the sum of the neutrino masses ${\sum m_{\nu}<0.72}$~eV from cosmology~\cite{Ade:2015xua}. The matrix $\mathcal{R}$ satisfies the constraint ${\mathcal{R}^{T}\mathcal{R}=1}$, which means that it can be parametrized as~\cite{Casas:2001sr}
\begin{equation}
 \mathcal{R}=
\begin{pmatrix}
 c_{12}c_{13} & s_{12}c_{13} & s_{13} \\
-s_{12}c_{23}-c_{12}s_{23}s_{13} & c_{12}c_{23}-s_{12}s_{23}s_{13} & s_{23}c_{13} \\
s_{12}s_{23}-c_{12}c_{23}s_{13} & -c_{12}s_{23}-s_{12}c_{23}s_{13} & c_{23}c_{13}
\end{pmatrix}
\label{equ:rmatrix}
\end{equation}
where ${c_{ij}=\cos(\omega_{ij})}$ and ${s_{ij}=\sin(\omega_{ij})}$ with $\omega_{ij}$ being a complex angle for which we choose ${\text{Re}(\omega_{ij}) \in [0,2\pi]}$ and ${\text{Im}(\omega_{ij}) \in [-8,8]}$. The dependence of the parameter $\text{Re}(\omega_{ij})$ is periodic~\cite{Drewes:2015iva}, whereas $\text{Im}(\omega_{ij})$ has no limit in general. Both statements can be verified by writing the sine and cosine in terms of the real and imaginary parts of the complex angle. We have constrained $\text{Im}(\omega_{ij})$, as a broader range is without consequence. Additionally, there are different sign conventions in the $\mathcal{R}$ matrix; however, this have no impact on the discussed observables such as the active-sterile mixing. We only consider the normal ordering, which affects the entries in $m_{\nu}^{\text{diag}}$. A normal hierarchical spectrum implies that  ${m_{1} \simeq 0}$, ${m_{2} \simeq \sqrt{\Delta m_{21}^{2}}}$ and ${m_{3} \simeq \sqrt{\Delta m_{32}^{2}}}$, whereas a degenerate spectrum with normal ordering implies that ${m_{\nu}^{\text{diag}}=m_{\text{min}}\text{diag}(1,1,1)}$ with ${\Delta m_{32}^{2}/m_{\text{min}} \ll 1}$. The masses of the sterile neutrinos are chosen from the interval ${M_{I} \in [0.1, 80]}$~GeV with the requirement ${M_{1} < M_{2} < M_{3}}$. We therefore will show the parameter space for the lightest ($M_1$) and heaviest ($M_3$) sterile neutrino separately, which means that our figures satisfy the paradigm ``one model, one dot''.

The Dirac mass matrix is obtained using \equ{casibapara}. Thereafter, experimental constraints on the physical observables are checked, such as the effective mass of neutrinoless double beta decay $m_{\beta \beta}$, the decay rate of the lepton flavor violating process $\mu \rightarrow e\gamma$, the active-sterile mixing, and the lifetime of the sterile neutrino $\tau_{N}$. If the experimental constraints are satisfied we keep the set of mass matrices ($Y_{\ell}, M_{D}$ and $M_{R}$); otherwise, we discard them. The method of calculating these observables and the experimental limits of them are taken from Ref.~\cite{Drewes:2015iva}. The constraints from direct searches and Big Bang nucleosynthesis (BBN) are especially important since these exclude parts of the parameter space of the active-sterile mixing; more detailed explanations will follow later.

Besides the Casas-Ibarra parametrization, we also generate random mass matrices
\begin{equation}
 M_{D} = m_{D}
\begin{pmatrix}
 c_{1} & c_{2} &  c_{3} \\
 c_{4} & c_{5} &  c_{6} \\
 c_{7} & c_{8} &  c_{9}
\end{pmatrix} \, , \hspace*{0.3cm}
 M_{R}= 
\begin{pmatrix}
 M_{1} &  0 & 0\\
 0 & M_{2} &  0\\
 0 &  0 & M_{3}
\end{pmatrix} \, ,
\end{equation}
where the charged lepton Yukawa matrix is diagonal, $M_{I}$ are the masses of the sterile neutrinos, $m_{D}$ controls the overall scale of the Dirac mass matrix, and $c_{j}$ are (independent) order 1 complex numbers with ${|c_{j}|=k_{j}}$ and ${\text{arg}(c_{j})=\phi_{j}}$ for ${j=\{1,...,9\}}$. We have chosen this structure of the mass matrices similar to the Casas-Ibarra parametrization. However, note that other possibilities with nondiagonal $M_{R}$ and $Y_{\ell}$ yield a similar result because the physical observables do not depend on the basis. We call this method the ``random case'' since the neutrino oscillation parameters are generated from the set of mass matrices randomly chosen in the flavor symmetry basis. This concept is similar in motivation but somewhat different in implementation from \textit{anarchy}, which postulates the independence of the measure~\cite{Brdar:2015jwo, Haba:2000be}. 

For this scenario with these mass matrices we use the ``generate-and-tune'' method to find viable realizations -- similar to the method in \Ref~\cite{Rodejohann:2015nva}:\footnote{Without such an approach, the hit rate for a viable realization would be very low.} we choose ${k_{j} \in [\epsilon, \frac{1}{\epsilon}]}$ and ${M_{I} \in [0.1, 80]}$ GeV randomly with the requirement ${M_{1} < M_{2} < M_{3}}$ and ${\epsilon=0.2}$ (the motivation for this quantity will be described below). Then, the phases $\phi_{j}$ and $m_{D}$ are picked to (locally) minimize the $\chi^{2}$-function, 
\begin{align}
 \chi^{2}&= \left(\frac{\theta_{12}-\theta_{12}^{\text{bf}}}{\sigma_{\theta_{12}}}\right)^2+\left(\frac{\theta_{13}-\theta_{13}^{\text{bf}}}{\sigma_{\theta_{13}}}\right)^2+\left(\frac{\theta_{23}-\theta_{23}^{\text{bf}}}{\sigma_{\theta_{23}}}\right)^2 \nonumber \\& +\left(\frac{\Delta m_{21}^{2}-(\Delta m_{21}^{2})^{\text{bf}}}{\sigma_{\Delta m_{21}^{2}}}\right)^2+\left(\frac{\Delta m_{32}^{2}-(\Delta m_{32}^{2})^{\text{bf}}}{\sigma_{\Delta m_{32}^{2}}}\right)^2 
\label{equ:chisquare}
\end{align}
where we use the best-fit values $({\theta_{ij}^{\text{bf}}}, (\Delta m_{ij}^{2})^{\text{bf}})$ and $1\sigma$ errors $({\sigma_{\theta_{ij}}, \sigma_{\Delta m_{ij}^{2}})}$ of the neutrino oscillation parameters from \Ref~\cite{GonzalezGarcia:2014bfa}. The minimization is performed with Brent's~\cite{Brentsmethod} and Powell's methods~\cite{Powell01011964}. Brent's method requires initial values for the $\phi_{j}$s and $m_{D}$ to perform the minimization. We choose $0$ and $2\pi$ for the $\phi_{j}$s and 10~keV and 300~keV for $m_{D}$.\footnote{These are the minimal and maximal values of $m_{D}$ corresponding to the sterile neutrino masses ${M_{I}\in[0.1,80]}$~GeV implying the scaling $m_{D} \sim \sqrt{M_{I}}$ from the seesaw mechanism.} If the final value of ${\chi^{2} < 9}$, ${\sum m_{\nu} < 0.72}$~eV and  the experimental constraints from \Ref~\cite{Drewes:2015iva} are satisfied, we keep the realization; otherwise, we discard it.

\subsection{Parameter space predictions versus experimental sensitivity}

\begin{figure*}[p!]
\centering
\begin{tabular}{c}
\includegraphics[scale=0.8]{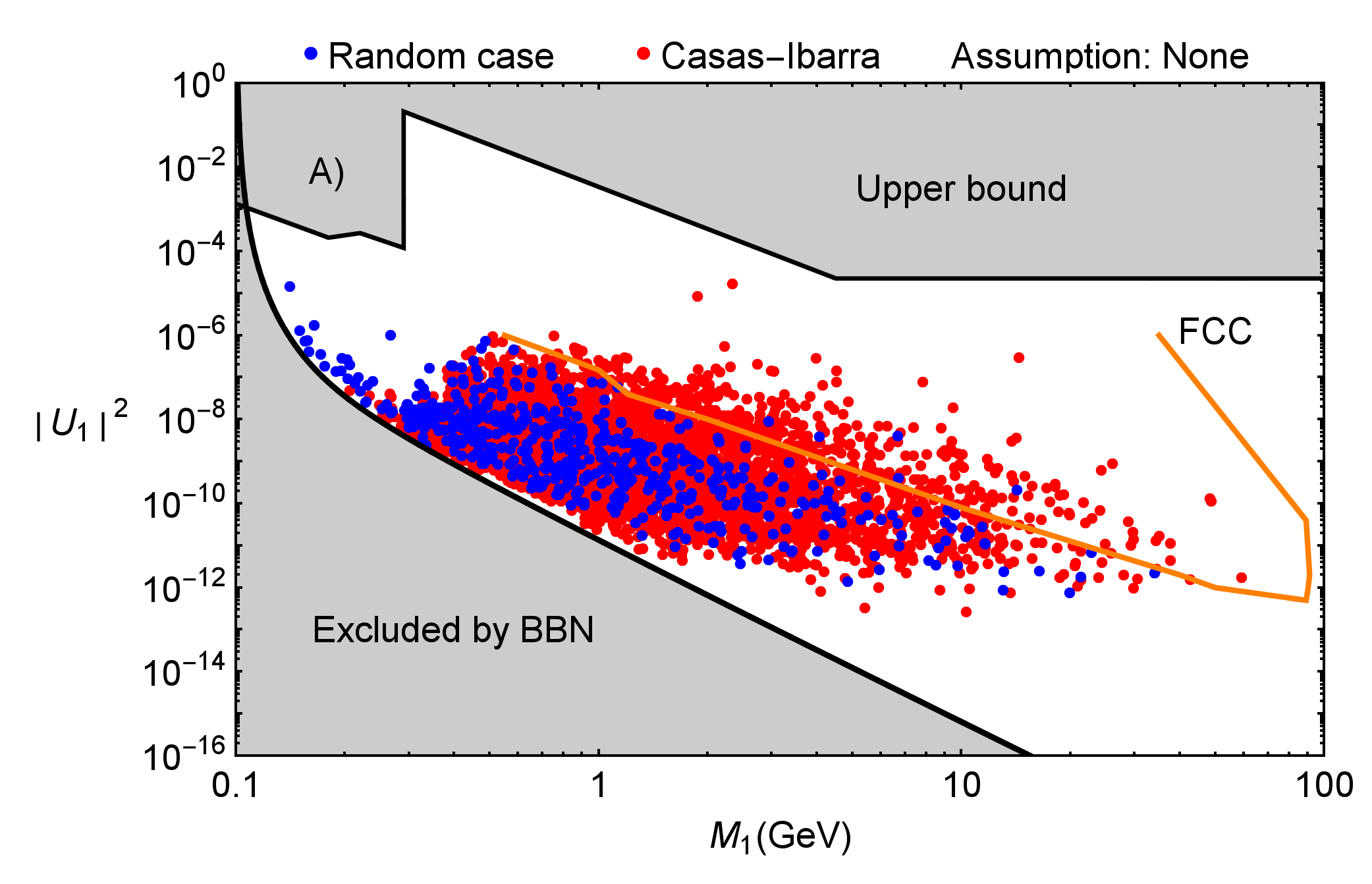} 
\\
\includegraphics[scale=0.8]{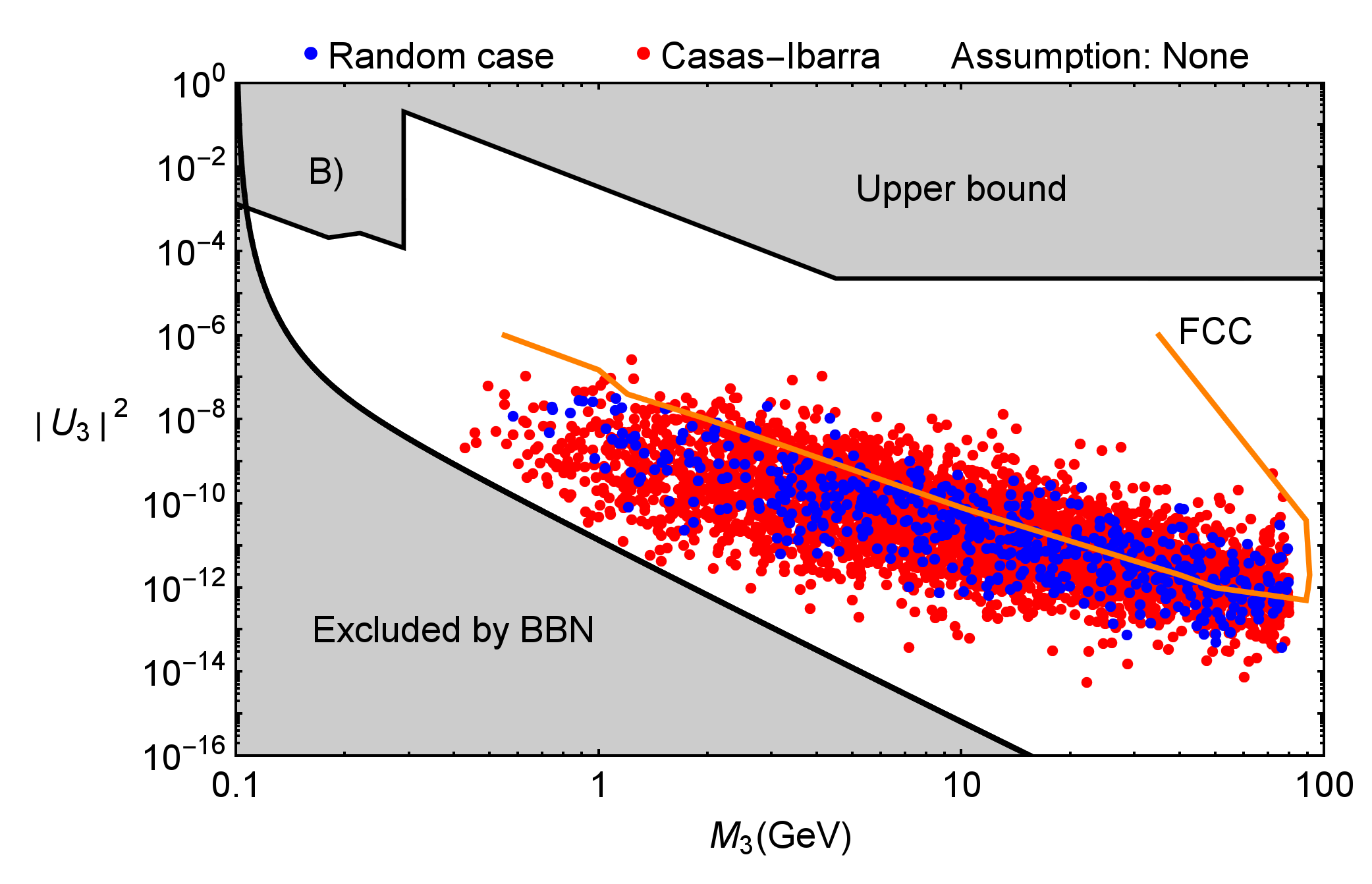}
\end{tabular}
\caption{Total active-sterile mixing predictions for the lightest (a) and heaviest (b) sterile neutrino for the Casas-Ibarra parametrization (red points) and random case (blue points), where one dot represents one model.  ``Assumption: None'' means that no assumption has been imposed on the ratio of mixings with the individual flavors. As a consequence, no information on SHiP and DUNE sensitivities can be given based on present information (which would either require sensitivity to mixing with the tau flavor or  direct sensitivity to the total mixing). We have taken the (optimistic) sensitivity bound expected from the FCC experiment~\cite{Blondel:2014bra}, which is directly sensitive to the total mixing. For the discussion of the bounds, see the main text; there is no seesaw bound because it does not apply to three generations of sterile neutrinos at the GeV scale.}
\label{fig:casibarand}
\end{figure*}

The realizations of the Casas-Ibarra parametrization and random case, which satisfy all experimental bounds, are shown in \figu{casibarand}, where the predictions for the total active-sterile mixing are plotted for the lightest and heaviest sterile neutrinos as a function of their mass.\footnote{Throughout the article, we omit our findings for the second-heaviest sterile neutrino since the parameter space of this particle is between the parameter space of the two other neutrinos because we enforce the requirement ${M_{1} < M_{2} < M_{3}}$.} In that figure, we compare the predictions obtained with the method outlined above with the sensitivity of future experiments. Since there is not yet any information on the sensitivities from SHiP and DUNE on the total mixing in the absence of any assumptions, we do not show the corresponding bounds.\footnote{The information from $|U_{eI}|^2$ and $|U_{\mu I}|^2$ can be, in principle, translated into the total mixing if it is known how much $|U_{\tau I}|^2$ contributes. In the absence of any assumption, there is no sensitivity from these elements as $|U_{\tau I}|^2$ is not measured. Direct information on the total mixing could come from processes such as ${N_{I} \rightarrow \eta \nu \rightarrow \pi^{+} \pi^{-} \pi^{0}+p^{\text{miss}}}$ but may be weaker than the bounds frequently shown in the literature.} The FCC experiment can directly measure the total mixing, as explained above.  

As it can be read off from the figure, both the random and  Casas-Ibarra cases tend to predict values in the same parameter space region. With the chosen procedure (varying the fundamental input parameters at random) the preferred parameter space is at the lower end of the allowed region, whereas FCC tests the upper section in terms of the mixing. Note, however, that in principle the whole shown (allowed) parameter space (\cf,~\cite{Drewes:2015iva}) can be reached, but larger total mixings require some fine-tuning. This can be shown using the Casas-Ibarra parametrization where the total mixing can be calculated analytically~\cite{Asaka:2015eda}
\begin{equation}
 |U_{I}|^{2}=\frac{1}{M_{I}}\sum_{j=1}^{3}m_{j}|\mathcal{R}_{jI}|^{2}
\label{equ:totalmix}
\end{equation}
where $M_{I}$ is the mass of the sterile neutrino $I$, $m_{j}$ is the mass of the light neutrino, and $\mathcal{R}_{jI}$ is the matrix element in \equ{rmatrix}. The $\mathcal{R}$ matrix depends on one complex angle $\omega_{ij}$ in the case with ${\mathcal{N}=2}$ sterile neutrinos, and it has been shown that the matrix element ${\mathcal{R}_{j I} \propto e^{|\text{Im}(\omega_{ij})|}}$ when ${|\text{Im}(\omega_{ij})| > 1}$~\cite{Alekhin:2015byh, Canetti:2012kh, Drewes:2015iva}. Therefore, having ${|\text{Im}(\omega_{ij})| \gg 1}$ leads, in general, to a large total mixing. In our case with ${\mathcal{N}=3}$ sterile neutrinos, the matrix elements behave similarly even though they depend on more than one complex angle. However, too large $|\text{Im}(\omega_{ij})|$ (either one or multiple angles) means that the mixing would violate the upper experimental bound. Therefore, they cannot be arbitrarily large and require some fine-tuning to probe the upper area of the parameter space at least within the method/parametrization chosen.

Let us now discuss the bounds shown in \figu{casibarand}, as  it is very important to compare sensitivities and bounds derived under similar assumptions. The lower bound comes from BBN; the observed abundances of light nuclei imply that the sterile neutrinos must have decayed long before BBN. This gives an upper bound on the lifetime of about $0.1$~s, and, consequently, a lower bound on the total mixing from the relationship ${\tau_{N} \sim \Gamma_{N}^{-1} \propto |U_{I}|^{-2}}$. Note that frequently a lower bound from the seesaw mechanism is shown. The scenario of ${\mathcal{N}=2}$ sterile neutrinos fixes this lower bound to~\cite{Alekhin:2015byh}
\begin{equation}
 |U_{I}|^{2} \gtrsim \frac{m_{\text{atm}}}{M_{N}} 
\begin{cases}
 \frac{m_{\odot}}{m_{\text{atm}}} & \text{Normal ordering (NO)} \\
  \frac{1}{2} & \text{Inverted ordering (IO)}
\end{cases}
\label{equ:lowboundtwocase}
\end{equation}
where ${m_{3} \simeq \sqrt{\Delta m_{31}^{2}}}>{m_{2} \simeq \sqrt{\Delta m_{21}^{2}}}>{m_{1} \simeq 0}$ for NO, ${m_{2} \simeq m_{1} \simeq \sqrt{|\Delta m_{31}^{2}|}}>{m_{3} \simeq 0}$ for IO, and $M_{N}$ is the overall scale of the sterile neutrino masses. Considering ${\mathcal{N}=3}$ sterile neutrinos, a lower bound can be derived using the Casas-Ibarra parametrization~\cite{Asaka:2015eda}
\begin{equation}
 |U_{I}|^{2} \geq \frac{m_{\text{min}}}{M_{I}}
\end{equation}
where $m_{\text{min}}$ is the mass of the lightest active neutrino. Since $m_{\text{min}}$ can be as low as zero, the seesaw bound is, in general, weaker than the BBN bound; therefore, we omit the seesaw bound. The upper bound comes from direct search experiments which constrain the active-sterile mixing by investigating different processes involving a sterile neutrino. A review  can be found in \Ref~\cite{Drewes:2015iva}, where the experimental upper limit on each individual mixing element is presented. The upper bounds on the individual mixing elements $|U_{e I}|^{2}$ and $|U_{\mu I}|^{2}$ are well constrained in the mass range 0.1-2 GeV with decreasing sensitivity for increasing mass, whereafter they reach a plateau at about 2-100 GeV. This is the exclusion limit reported by the DELPHI Collaboration~\cite{DELPHI} which were sensitive to this mass range. The mixing element $|U_{\tau I}|^{2}$ is best constrained in the mass range 2-100 GeV by the same plateau mentioned before, whereas it is less constrained below 2 GeV because of the tau production threshold. Therefore, a different method of identifying the flavor of the light neutrino has to be used   below 2 GeV (which is usually done by identifying the associated lepton). As a consequence, the constraint on the total mixing in \figu{casibarand} is typically limited by the sensitivity to $|U_{\tau I}|^{2}$. 

\begin{figure*}[p!]
\centering
\begin{tabular}{c}
\includegraphics[scale=0.8]{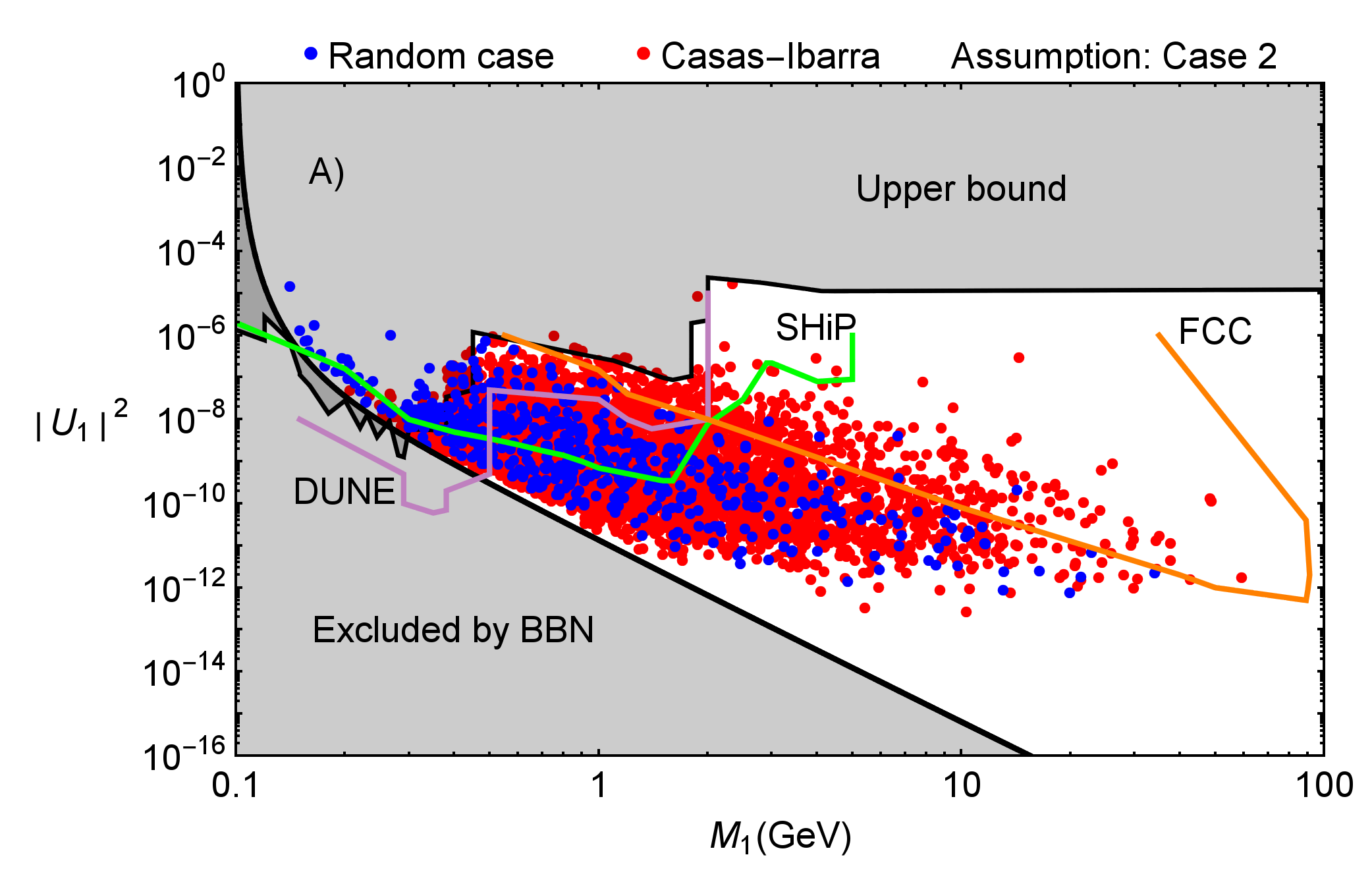} 
\\
\includegraphics[scale=0.8]{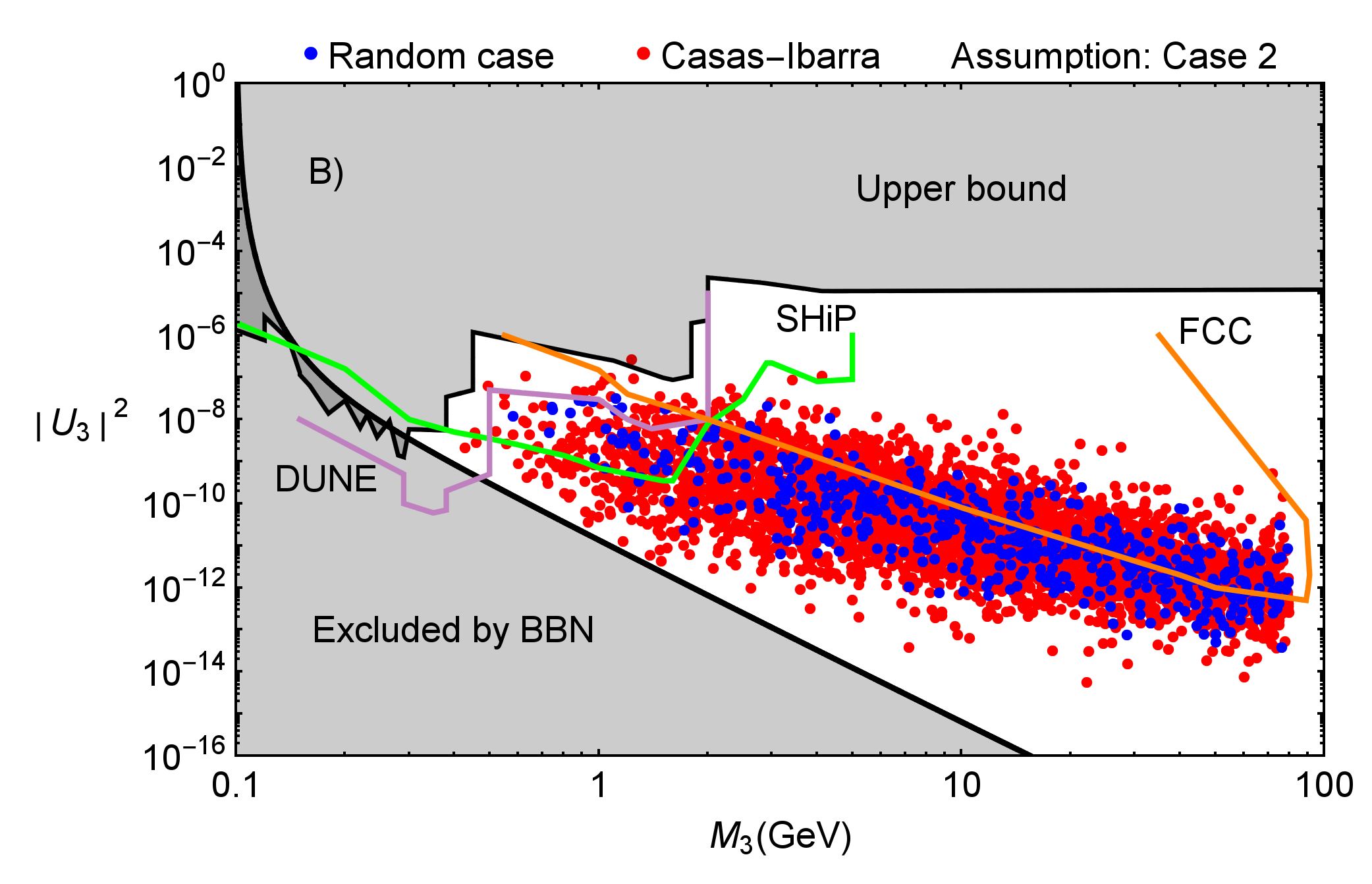}
\end{tabular}
\caption{
Same as \figu{casibarand} with the lightest (a) and heaviest (b) sterile neutrino, where the experimental bounds are shown for the assumption ${|U_{e I}|^{2}:|U_{\mu I}|^{2}:|U_{\tau I}|^{2}=1:16:3.8}$ [this is indicated by the statement ``Assumption: Case~2'' \equ{assumption}]. We have taken the optimistic sensitivity bounds from FCC~\cite{Blondel:2014bra} and DUNE~\cite{Adams:2013qkq}, whereas we have taken the case~2 [\equ{assumption}] scenario as a sensitivity bound for the SHiP experiment~\cite{Anelli:2015pba}. }
\label{fig:casibarandassumptions}
\end{figure*}

Better constraints from the experiments can be typically obtained if one assumes specific relationships among the $|U_{\alpha I}|^{2}$. In that case, the sensitivity to the total mixing is not necessarily constrained by $|U_{\tau I}|^{2}$ if the mixing with the tau element is small enough. For illustration, we use the assumption of case~2 [\equ{assumption}] in the following, which means that the sensitivity to the total mixing will dominated by $|U_{\mu I}|^{2}$.\footnote{The contribution from each flavor to the upper bound on the total mixing is calculable by using the ratio in case~2 [\equ{assumption}],
\begin{equation}
 {\frac{|U_{e I}|^{2}}{|U_{I}|^{2}}:\frac{|U_{\mu I}|^{2}}{|U_{I}|^{2}}:\frac{|U_{\tau I}|^{2}}{|U_{I}|^{2}}=0.05:0.77:0.18}.
\end{equation}
Since the electron and muon flavors are constrained much better than the tau flavor, they will typically dominate when translating the individual bounds to the bound on the total mixing even if the sterile neutrino mixes substantially with the tau flavor. Only if the sterile neutrino mixes {\em only} with the tau flavor, the upper bound will come directly from the upper bound on the tau mixing element.}

We show in \figu{casibarandassumptions} the bounds using this assumption in relationship to the predictions for the active-sterile mixings. Note that our predictions generally follow the trend of the upper bound (because the models are constrained by the individual mixings implied there), but they are not produced using this ratio of the active-sterile mixings and can therefore violate it. Most importantly, bounds from the SHiP and DUNE experiments can be included now, as the individual mixing sensitivities can be translated into the total mixing sensitivity. The upper bound and the bounds of these experiments have been derived using the same method; the sensitivities of the future experiments should be interpreted with respect to these bounds. For example, SHiP can improve the current bound on the total mixing by about 2 orders of magnitude in the energy range ${M \lesssim 2}$~GeV. Note that the lower (BBN) bound does not depend on the assumptions for the active-sterile mixing ratios, because the sterile neutrino lifetime depends on the total mixing only; see above. 

\section{Parameter space from flavor models}

Here, we assume that the structure of the mass matrices comes from a flavor symmetry, and we discuss the implications of that assumption. 

\subsection{Method and flavor symmetry model}

\begin{table*}
\makebox[\textwidth]{
\begin{tabular}{c|c|c|c|c}
 \#\ & $Y_{\ell} = M_{\ell}/v$ & $M_{D}/m_{D}$ & $M_{R}/m_{R}$ & $G_{F}$ \\
\hline \\[-2.5mm]
15 & $ \vspace{1mm} \hspace{1mm}\left(\hspace{-1mm}\ba{ccc}
\epsilon^{4}&\epsilon^{4}&\epsilon^{2} \\ \epsilon^{3}&\epsilon^{4} & 1 \\ \epsilon^{3}&\epsilon^{2} & 1 \ea\hspace{-1mm}\right)$ & $
 \hspace{1mm}\left(\hspace{-1mm}\ba{ccc}
\epsilon^{2}&\epsilon &\epsilon^{3} \\ \epsilon^{2}&\epsilon &\epsilon^{2} \\ \epsilon &\epsilon^{2} & 1 \ea\hspace{-1mm} \right)$ & $
\hspace{1mm}\left(\hspace{-1mm}\ba{ccc}
\epsilon^{2}&\epsilon^{2}&\epsilon \\ \epsilon^{2}&\epsilon &\epsilon^{2} \\ \epsilon &\epsilon^{2} & 1 \ea\hspace{-1mm}\right)$ & $Z_{5}\times Z_{7}$ \\
\hline \\[-2.5mm]
19 & $ \vspace{1mm}
\hspace{1mm}\left(\hspace{-1mm}\ba{ccc}
\epsilon^{4}&\epsilon^{4}&\epsilon^{2} \\ \epsilon^{2}&\epsilon^{2}&\epsilon^{2} \\ \epsilon^{4}&\epsilon^{2} & 1 \ea\hspace{-1mm}\right)$ & $

\hspace{1mm}\left(\hspace{-1mm}\ba{ccc}
\epsilon &\epsilon^{2}&\epsilon \\ \epsilon &  1 & \epsilon \\ \epsilon &  1 & \epsilon \ea\hspace{-1mm}\right)$ & $
\hspace{1mm}\left(\hspace{-1mm}\ba{ccc}
\epsilon &\epsilon^{2}&\epsilon^{5} \\ \epsilon^{2} & 1 & \epsilon^{3} \\ \epsilon^{5}&\epsilon^{3} & 1 \ea\hspace{-1mm}\right)$ & $Z_{5}\times Z_{6}$ \\
\hline \\[-2.5mm]
22 & $ \vspace{1mm}
\hspace{1mm}\left(\hspace{-1mm}\ba{ccc}
\epsilon^{4}&\epsilon^{3}&\epsilon^{2} \\ \epsilon^{2}&\epsilon^{2}&\epsilon^{3} \\ \epsilon^{5}&\epsilon &  1 \ea\hspace{-1mm}\right)$ & $

\hspace{1mm}\left(\hspace{-1mm}\ba{ccc}
\epsilon^{2}&\epsilon &\epsilon^{2} \\ 1&\epsilon &  1 \\ 1&\epsilon^{3} & 1 \ea\hspace{-1mm}\right)$ & $
\hspace{1mm}\left(\hspace{-1mm}\ba{ccc}
1&\epsilon^{3} & 1 \\ \epsilon^{3}&\epsilon &\epsilon^{3} \\ 1&\epsilon^{3} & 1 \ea\hspace{-1mm}\right)$ & $Z_{3}\times Z_{9}$ \\ \hline
\end{tabular}}
\caption{Selected examples for texture sets $Y_\ell$, $M_D$, and $M_R$ from flavor models~\cite{Plentinger:2008up}, where the numbering of each texture set is kept from the original article. The last column shows the flavor symmetry extension of the SM symmetry, \ie, ${G_{\text{SM}} \times G_{F} = SU(3) \times SU(2) \times U(1) \times G_{F}}$, that will realize the structure of the matrices.}
\label{tab:texture}
\end{table*}

In order to illustrate the impact of flavor symmetry models, we start off from sets of textures shown in \Tab~\ref{tab:texture} from \Ref~\cite{Plentinger:2007px,Plentinger:2008up}, which can be obtained from the discrete product flavor symmetry groups shown in the last column.\footnote{Note that we have in fact checked all examples from \Ref~\cite{Plentinger:2008up} but only show a few examples here for illustration.} These textures represent the leading order structure of the mass matrix elements, normalized such that the largest element is order unity. The original motivation to derive these textures has been to describe all masses and mixings with a single parameter ${\epsilon \simeq \theta_C \simeq 0.2}$, which may be a remnant of a grand unified theory -- which is a concept introduced as ``extended quark-lepton complementarity''. For example, it is well known that the quark masses and mixings can be approximated by powers of $\epsilon$, such as the famous Wolfenstein parametrization~\cite{PhysRevLett.51.1945} or the quark and charged lepton masses ${m_u:m_c:m_t \sim \epsilon^6:\epsilon^4:1}$, ${m_d:m_s:m_b \sim \epsilon^4:\epsilon^2:1}$, and ${m_e:m_\mu:m_\tau \sim \epsilon^4:\epsilon^2:1}$. Similarly, the neutrinos can, for the normal hierarchy, be described by ${m_1:m_2:m_3 \sim \epsilon^2:\epsilon:1}$. Assuming that the lepton mixings can be described by powers of $\epsilon$ or maximal mixings (which could come from an additional symmetry) as well, one can list the set of textures which can describe two large lepton mixing angles and a small $\theta_{13}$; see \Ref~\cite{Plentinger:2006nb} for details of the method.

The texture of each mass matrix is obtained by assigning charges to the leptons under the flavor symmetry $G_{F}=Z_{n_{1}} \times Z_{n_{2}} \times \cdot \cdot \cdot \times Z_{n_{m}}$~\cite{Plentinger:2008up}, namely
\begin{align}
 (e_{R})_{i} &\sim (p^{i}_{1}, p^{i}_{2},...,p^{i}_{m})=\boldsymbol{p}^{i}, \\ \ell_{i} &\sim(q^{i}_{1}, q^{i}_{2},...,q^{i}_{m})=\boldsymbol{q}^{i}, \\ (N_{R})_{i} & \sim (r^{i}_{1}, r^{i}_{2},...,r^{i}_{m})=\boldsymbol{r}^{i}
\end{align}
for the right-handed lepton $(e_{R})_{i}$, the lepton doublet $\ell_{i}$, and the right-handed neutrino $(N_{R})_{i}$, respectively. The $j$th entry in each row vector denotes the $Z_{n_{j}}$ charge of the particle, $i=1,2,3$ is the generation index, $m$ is the number of $Z_{n}$ factors and $n_{k} \ (k=1,2,...,m)$ may be different. 

In the Froggatt-Nielsen (FN) framework~\cite{Froggatt1979277}, there exists a scalar flavon field $f_{n_{k}}$ for each of the $Z_{n_{k}}$. Each flavon is only charged under its associated $Z_{n_{k}}$ factor, whereas it is a singlet under the SM flavor symmetries and all other $Z_{n_{j}}$ with $j \neq k$. Each flavon acquires a nonzero universal vacuum expectation value $\langle f_{n_{k}} \rangle \simeq v_{f}$ that spontaneously breaks the $Z_{n_{k}}$ factor. Additionally, beside coupling to the SM Higgs, the leptons also couple to superheavy fermions with universal mass $M_{F}$, and integrating them out leads to the hierarchical structure in the Yukawa/mass matrices with $\epsilon \simeq v_{f}/M_{F} \simeq 0.2$ being the same order as the Cabibbo angle. Therefore, the lepton mass terms in the FN framework becomes
\begin{align}
\mathcal{L}_{Y} = & -(\Pi_{k=1}^{m}\epsilon^{\alpha_{ij}^{k}})x_{ij}H^{*}\ell_{i}(e_{R})_{j} \nonumber \\
 & -(\Pi_{k=1}^{m}\epsilon^{\beta^{k}_{ij}})y_{ij}\tilde{H}\ell_{i}(N_{R})_{j} \nonumber \\
& - \frac{1}{2}m_{R}(\Pi_{k=1}^{m}\epsilon^{\gamma^{k}_{ij}})z_{ij}(N_{R})_{i}(N_{R}^{c})_{j} +h.c. \nonumber
\end{align}
where $x$, $y$, and $z$ are independent order unity complex numbers and $\tilde{H}=i\sigma^{2}H$. This leads to effective SM lepton masses that are suppressed by $\epsilon$ where the exponent is determined by the quantum numbers of the leptons~\cite{Plentinger:2008up}, 
\begin{align}
 \alpha_{ij}^{k} & =\text{min}[(p_{i}^{k}+q_{j}^{k}) \ \text{mod} \ n_{k}, (-p_{i}^{k}-q_{j}^{k}) \ \text{mod} \ n_{k}], \nonumber \\
 \beta_{ij}^{k} & =\text{min}[(q_{i}^{k}+r_{j}^{k}) \ \text{mod} \ n_{k}, (-q_{i}^{k}-r_{j}^{k}) \ \text{mod} \ n_{k}], \nonumber \\
 \gamma_{ij}^{k} & =\text{min}[(r_{i}^{k}+r_{j}^{k}) \ \text{mod} \ n_{k}, (-r_{i}^{k}-r_{j}^{k}) \ \text{mod} \ n_{k}]. \nonumber
\end{align} 

Therefore, the texture arise as the leading order products of $\epsilon$ for a certain Yukawa coupling or mass matrix. The lepton mass matrix elements are therefore the given by 
\begin{align}
 (M_{\ell})_{ij}  \approx v \Pi_{k=1}^{m} \epsilon^{\alpha_{ij}^{k}}x_{ij}, &\hspace*{0.5cm}
 (M_{D})_{ij} \approx m_{D} \Pi_{k=1}^{m} \epsilon^{\beta_{ij}^{k}}y_{ij},  \nonumber \\
 (M_{R})_{ij} & \approx m_{R} \Pi_{k=1}^{m} \epsilon^{\gamma_{ij}^{k}}z_{ij}, \nonumber
\end{align} 
where $v$ is the Higgs vacuum expectation value and $m_{D} \ (m_{R})$ is the overall scale of the Dirac (Majorana) mass matrix. 

Here, we reinterpret the textures from \Ref~\cite{Plentinger:2007px} obtained for the seesaw mechanism, which can be described by flavor symmetries~\cite{Plentinger:2008up}, in terms of the FN framework~\cite{Froggatt1979277}. Note that the original textures were produced using  ${\theta_{13} \equiv 0}$, whereas more recent experimental data show ${\theta_{13} \neq 0}$~\cite{Abe:2013sxa, An:2012eh, Ahn:2012nd}. Here, we apply the FN mechanism literally, which means that each entry in the mass matrices can be modified by an independent order 1 complex number $c_{ij}$ with ${|c_{ij}|=k_{ij}}$ and ${\text{arg}(c_{ij})=\phi_{ij}}$ (before we used $x, y$, and $z$ as complex numbers). We need 24 order 1 complex numbers when studying the texture sets; the charged lepton Yukawa matrix and the Dirac mass matrix needs nine order 1 complex numbers each (one for each element), whereas the Majorana mass matrix only needs six since some of the matrix elements are not independent due to the constraint ${M_{R}=M_{R}^{T}}$. The Majorana mass matrix has to obey this requirement because of the symmetric mass term in \equ{massseesaw}. As a consequence, some realizations of the texture sets will satisfy observations with a nonzero value of $\theta_{13}$. The correspondingly predicted parameter space for the HNLs will then be a direct consequence of the flavor symmetry. 

We use the ``generate-and-tune'' method introduced in the previous section to generate viable realizations, where we choose the overall scale of the Majorana mass matrix ${m_{R} \in [0.1, 100]}$ GeV and the absolute value of the order 1 complex number ${k_{j} \in [\epsilon, \frac{1}{\epsilon}]}$ and fix ${\epsilon=0.2}$. The initial values for the $\phi_{j}$s are the same as they were previously. However, we cannot easily normalize the overall scale of the Dirac mass matrix $m_{D}$ to the mass of the sterile neutrino due to the nondiagonal Majorana mass matrix $M_{R}$.  Therefore, we choose the starting values of the minimizer $0.5\sqrt{m_{R}}$ and $1.5\sqrt{m_{R}}$, where the coefficients give more freedom to the minimization compared to fixing $m_{D}=\sqrt{m_{R}}$. Since we cannot guarantee the masses of the sterile neutrinos to be  in the interval ${M_{I} \in [0.1, 80]}$ GeV, which is of interest to us, we can use the rescaling
\begin{equation}
 M_{R} \rightarrow zM_{R} \hspace*{0.5 cm} \text{and} \hspace*{0.5 cm} M_{D} \rightarrow \sqrt{z}M_{D}
\end{equation}
for a real number $z$ if one (or multiple) masses are outside the interval ${M_{I} \in [0.1, 80]}$~GeV.

\subsection{Parameter space predictions versus experimental sensitivity}

\begin{figure*}[p!]
\centering
\begin{tabular}{c}
\includegraphics[scale=0.8]{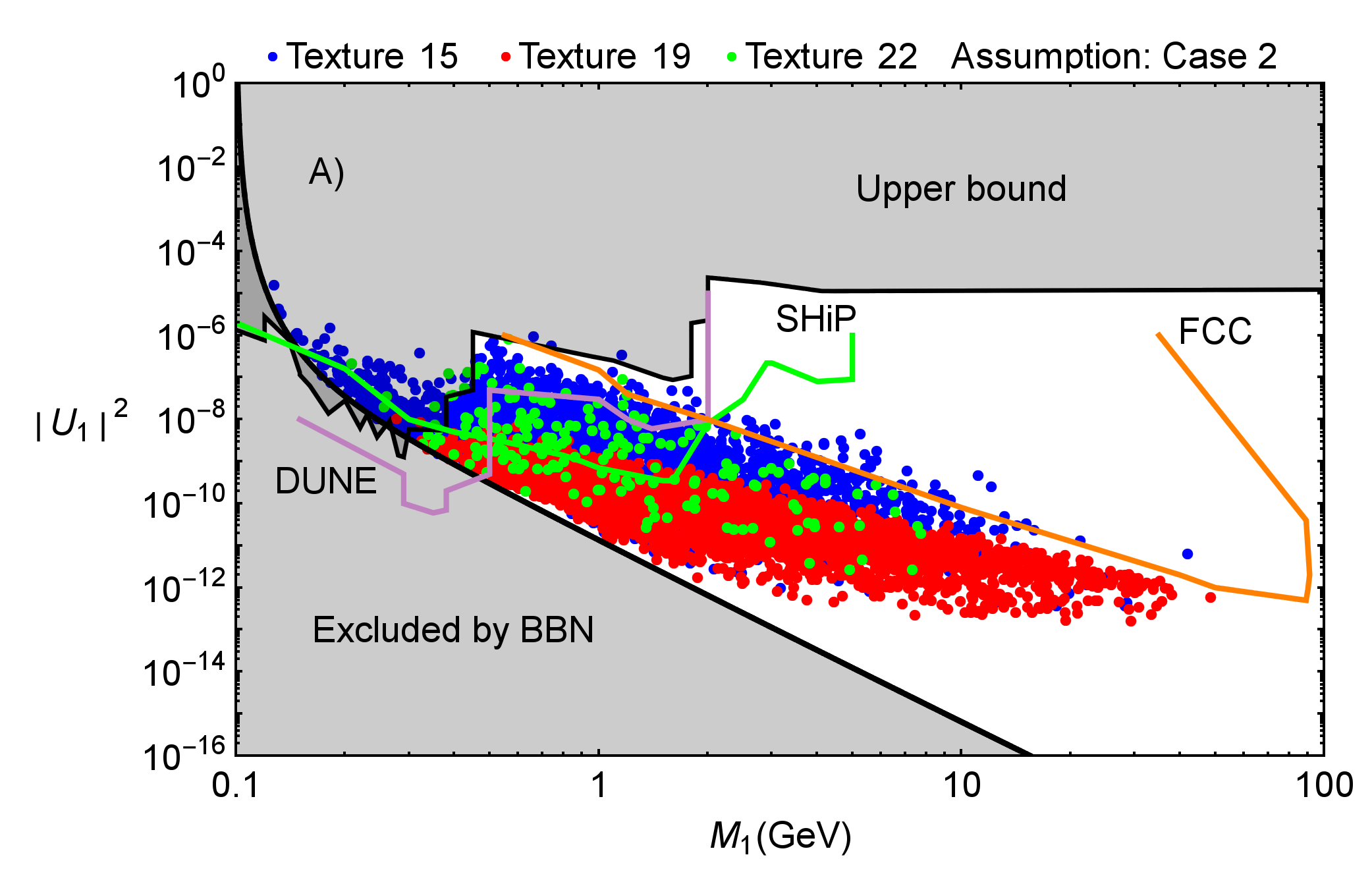}
\\
\includegraphics[scale=0.8]{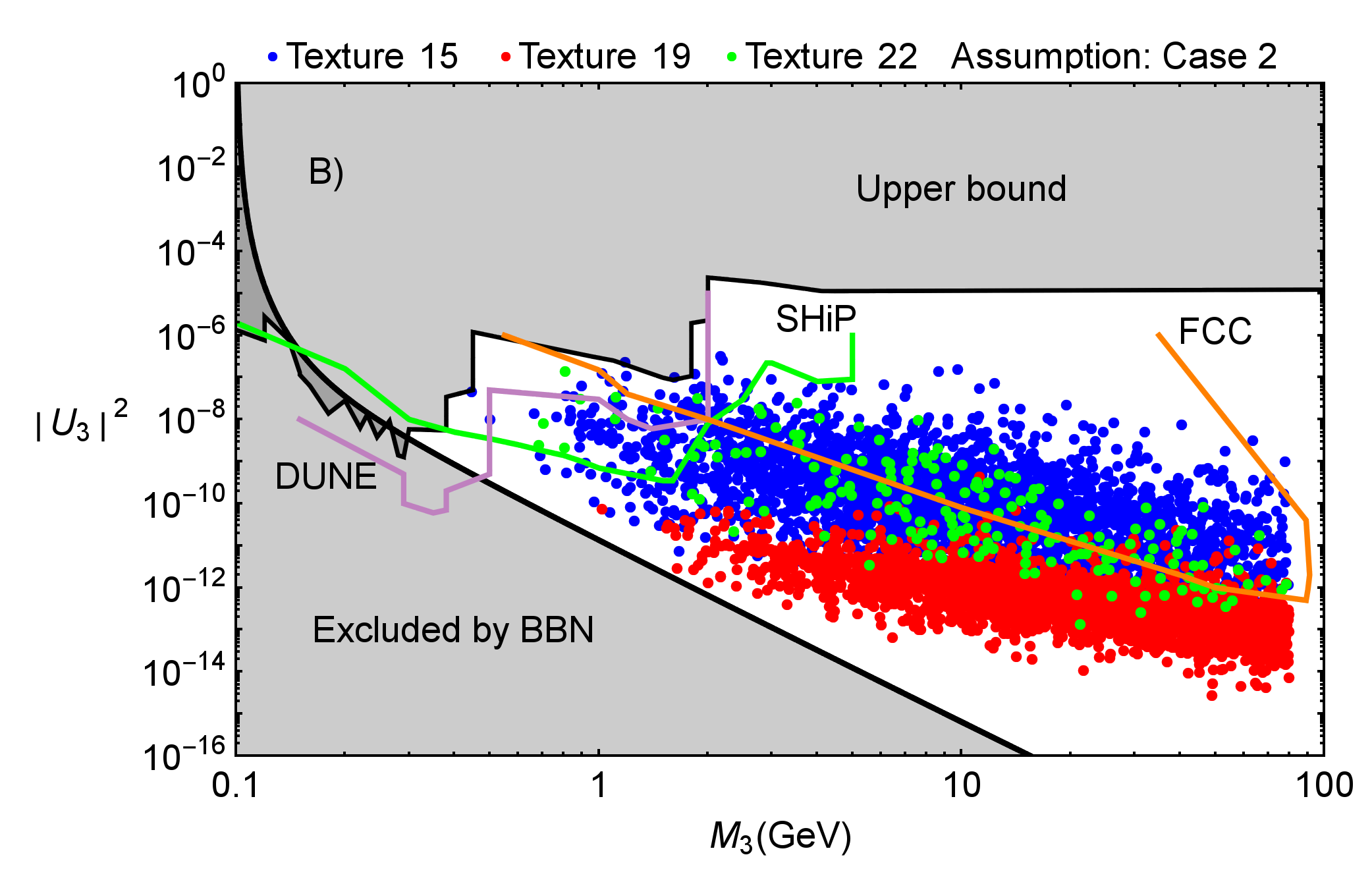}
\end{tabular}
\caption{Figure similar to \figu{casibarandassumptions} for the predictions of the total active-sterile mixing of the lightest (a) and heaviest (b) sterile neutrino from different texture sets (colors); see the figure legend. The assumption ${|U_{e I}|^{2}:|U_{\mu I}|^{2}:|U_{\tau I}|^{2}=1:16:3.8}$ has been included for sensitivities and experimental bounds.}
\label{fig:totalmix}
\end{figure*}

A comparison among the parameter space predictions for the total mixings from different texture sets is shown in \figu{totalmix}. One can read off from that figure that the flavor symmetry controls the size of the total mixing with the  sterile neutrinos in some (not all) cases. For example, texture~19 produces small mixings beyond the reach of future experiments, whereas texture~22 produces larger mixings which tend to be in the reach of SHiP. Texture~15 occupies a large parameter space, where the predictions tend to contain many models with large mixings. As a consequence, future measurements of HNL can be used as a model discriminator. However, certain textures cannot be distinguished based on the total mixing only, \eg, textures~15 and~22, as seen in \figu{totalmix}. 
We therefore study the flavor-dependent measurements in the next subsection.

\begin{figure*}[p!]
\centering
\begin{tabular}{c}
\includegraphics[scale=0.8]{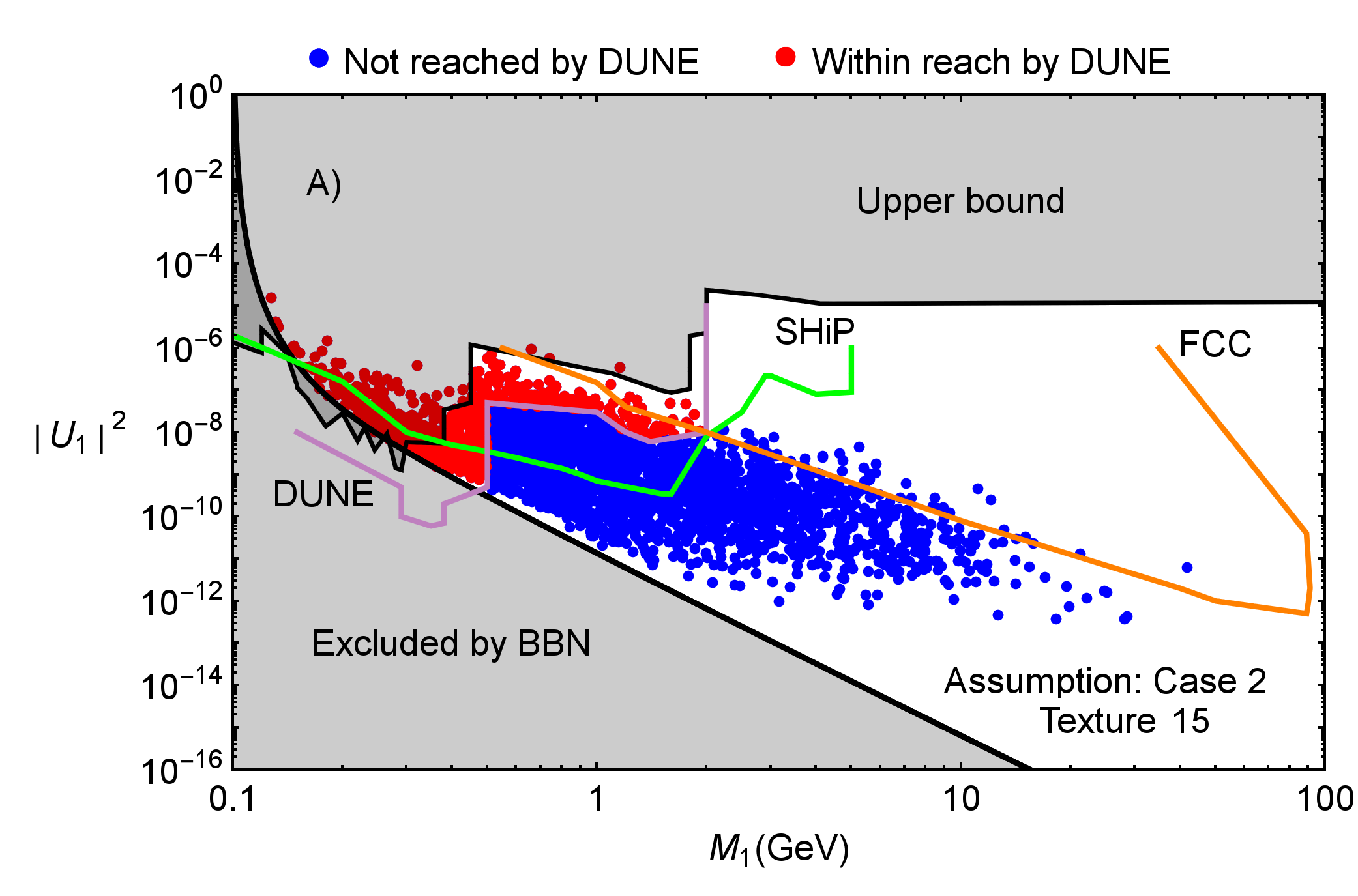}
\\
\includegraphics[scale=0.8]{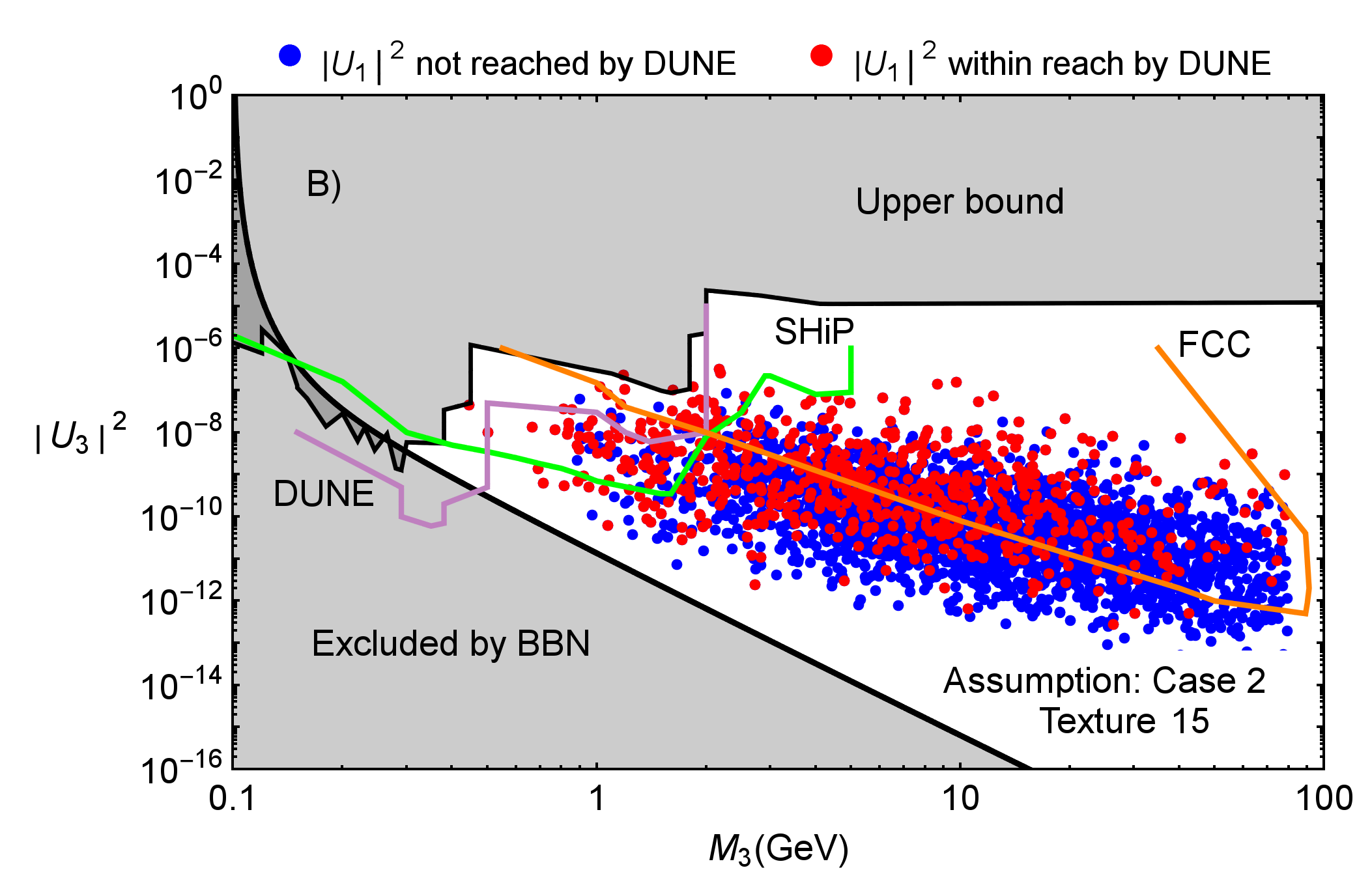}
\end{tabular}
\caption{Similar to \figu{totalmix} for texture 15 only where we focus on the complementary among the different experiments. The red (blue) points are the model predictions with the mixing $|U_{1}|^{2}$ (not) reachable by the DUNE experiment as is obvious from subfigure (a). In subfigure (b), the corresponding mixing for the heaviest state is shown for the same models; some of the red points are within reach of other experiments, but others are not. This means that DUNE can exclude model predictions for the heaviest sterile neutrino which are out of reach by FCC, and vice versa.}
\label{fig:comp}
\end{figure*}
\vspace{-0.1cm}
Each model predicts $\mathcal{N}=3$ sterile neutrinos at the GeV scale which can be sought experimentally. Each experiment has the potential to discover either of them; however, they are also complementary to each other when probing the parameter space and excluding model predictions. We use texture 15 when investigating the complementary among the experiments -- however, it can be done for every texture set. The model predictions for the lightest sterile neutrino within reach of the DUNE experiment are shown as red points in \figu{comp}(A) whereas the blue points are not within reach by DUNE. This leads to two subsets of the model predictions for texture 15. In comparison to the lightest sterile neutrino, we also show the heaviest sterile neutrino for the two subsets in \figu{comp}(B). The coloring of these points depends on whether or not the total mixing of the lightest sterile neutrino in the model is within reach of the DUNE experiment. The red (blue) points mean the lightest sterile neutrino can(not) be probed by the DUNE experiment. Some of the red points in \figu{comp}(B) are already within reach of the other experiments; however, others are not. Therefore, DUNE is complementary in excluding model predictions of the heaviest sterile neutrino by investigating the parameter space of the lightest sterile neutrino. Including additional parts of the parameter space probed by the other experiments means a larger fraction of the blue points would become red. Note that we have not considered the second-lightest sterile neutrino in this context; adding that would simply exclude more model predictions. A similar discussion could also be done for the model predictions probed by FCC or SHiP and the different sterile neutrinos. Combining all bounds from DUNE, SHiP and FCC gives the strongest upper bound; however, there are still cases which cannot be excluded even in this situation. To probe these, an experiment with the production of the sterile neutrinos from b-mesons is needed \cite{RichardJacobsson}. 

\vspace{-0.25cm}

\section{Flavor-dependent measurements}

\begin{figure*}[p!]
\centering
\begin{tabular}{c}
\includegraphics[scale=0.8]{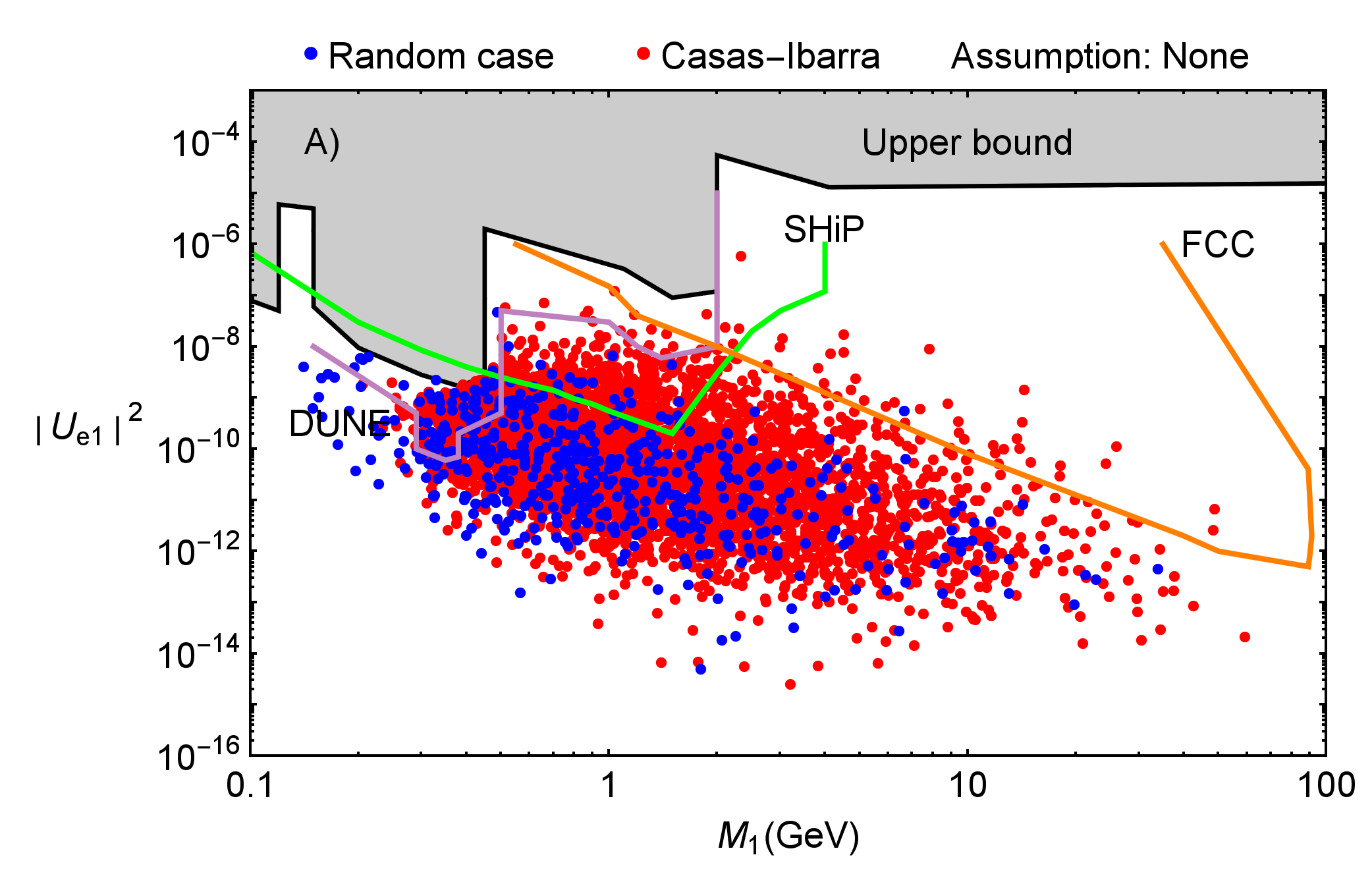}
\\
\includegraphics[scale=0.8]{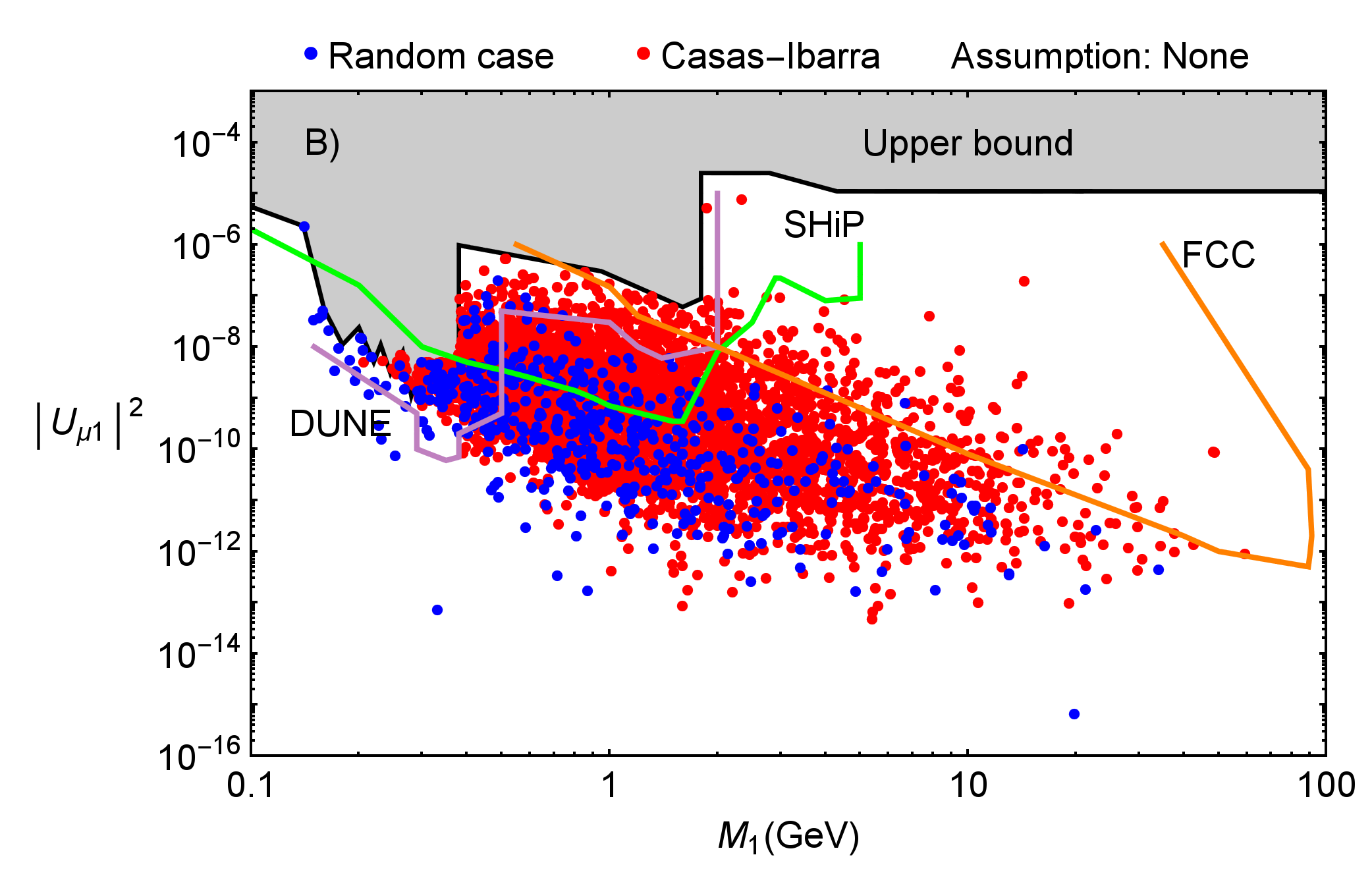}
\end{tabular}
\caption{The individual mixing elements $|U_{e 1}|^{2}$ (a) and $|U_{\mu 1}|^{2}$ (b) for the lightest sterile neutrino for the Casas-Ibarra parametrization and random cases. The upper bound from direct search experiments is shown as well, whereas no lower bound exists for three generations of sterile neutrinos (since one mixing element can be very small if another is large to ensure the lifetime bound of the sterile neutrino relevant for BBN). The bounds have been translated from \Refs~\cite{Anelli:2015pba,Adams:2013qkq,Blondel:2014bra}.}
\label{fig:individualcasibarand}
\end{figure*}

Here we investigate the model predictions vs experimental bounds for the individual mixing elements $|U_{e I}|^{2}$ and $|U_{\mu I}|^{2}$ the future experiments SHiP and DUNE are primarily sensitive to.\footnote{The tau mixing element $|U_{\tau I}|^{2}$ is not shown since the main source of sterile neutrinos for the SHiP and DUNE experiments are from charmed hadrons which have a similar mass of the tau-lepton -- meaning $|U_{\tau I}|^{2}$ may only contribute to production (via decays of tau-leptons from $D_{s}$-mesons) because of the mass difference between the charm meson and the tau-lepton. Therefore, it is considered irrelevant for subsequent sterile neutrino decays~\cite{Gorbunov:2013dta, Heikolacker}.} 
 
Our results for the random and Casas-Ibarra cases are shown in \figu{individualcasibarand}. Let us discuss the current and future experimental bounds first. The upper bound comes from direct search experiments, which, as we have discussed earlier, are directly sensitive to the depicted mixing elements, as are the future experiments SHiP and DUNE. There is no lower bound since one mixing element can be very small if another is large to ensure the lifetime constraint on the sterile neutrino ${\tau_{N}<0.1}$~s. While the FCC experiment is insensitive to the individual mixings, the individual mixings have to satisfy the constraint on the total mixing; therefore, the FCC bound applies here.  In summary, no assumptions on the ratios of the active-sterile mixings have been included in the exclusion bounds indicated by the statement ``Assumption: None'' in the top of the figure.

Regarding the model predictions, the Casas-Ibarra parametrization occupies more of the parameter space than the random case because the neutrino oscillation parameters are used as input rather than as a constraint, whereas the random model requires some fine-tuning to obtain large mixings. Note, however, that the Casas-Ibarra parametrization does not represent any model prediction and that the Casas-Ibarra and random cases can in principle reproduce the same region of the parameter space. Comparing with the sensitivities from SHiP and DUNE, large fractions of the parameter space can be probed, where $|U_{\mu I}|^{2}$ tends to have a better sensitivity. It is noteworthy that in spite of a missing lower bound, extremely small mixings are rarely predicted as well. Extremely small mixings might require some fine-tuning such as cancellations among the different terms in the active-sterile mixing matrix \equ{activesterilemixing}. The FCC experiment can constrain a small part of the realizations here -- however, FCC is intended to search for heavier sterile neutrinos (we show these plots only for $M_1$). The FCC bound is weaker in constraining the individual mixing elements compared to constraining the total mixing since ${|U_{I}|^{2} \geq |U_{\alpha I}|^{2}}$. Note that here no realizations are above the upper bound, since it directly applies here (without assumptions, the models may not satisfy).

\begin{figure*}[p!]
\centering
\begin{tabular}{c}
\includegraphics[scale=0.8]{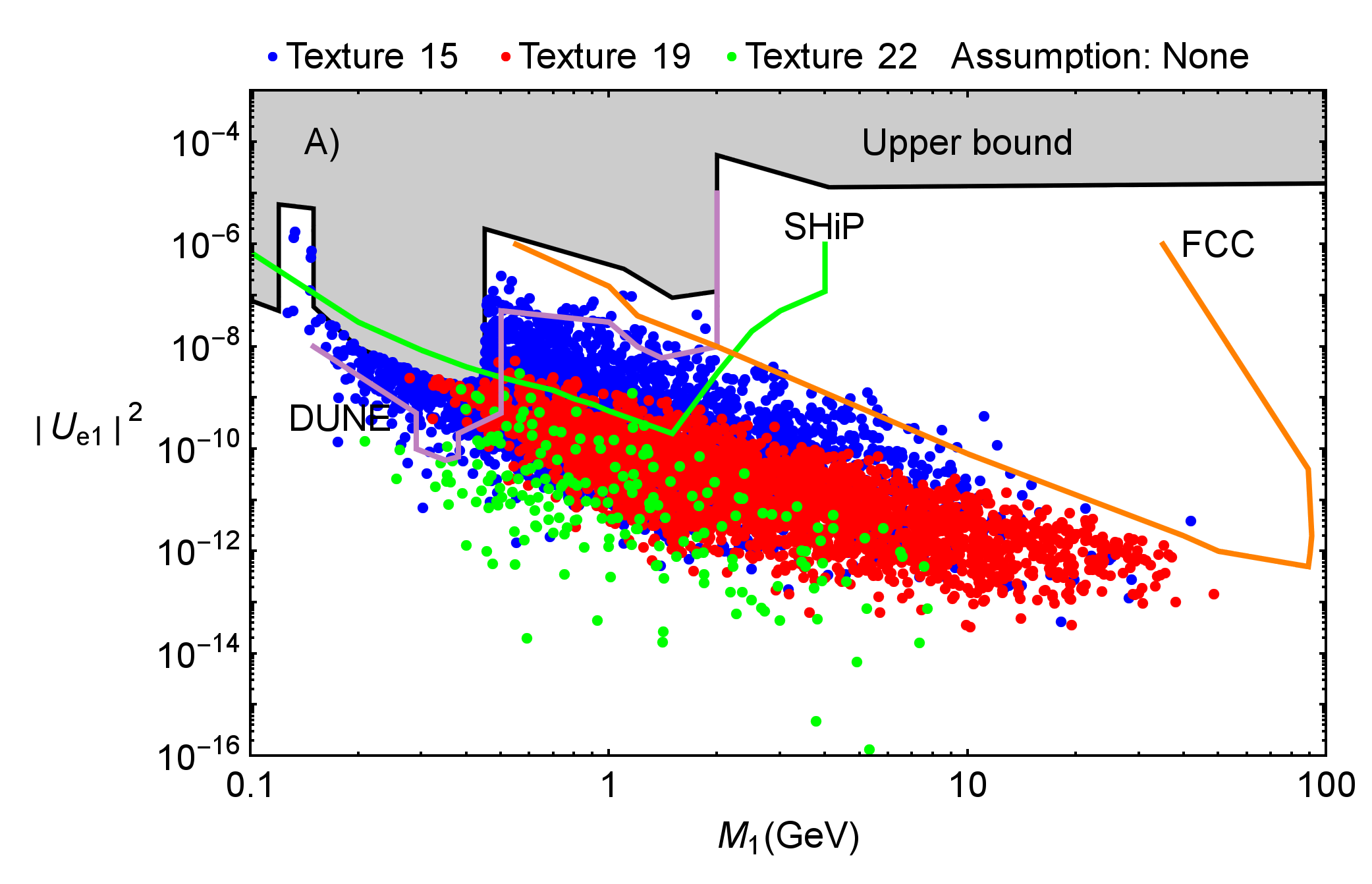}
\\
\includegraphics[scale=0.8]{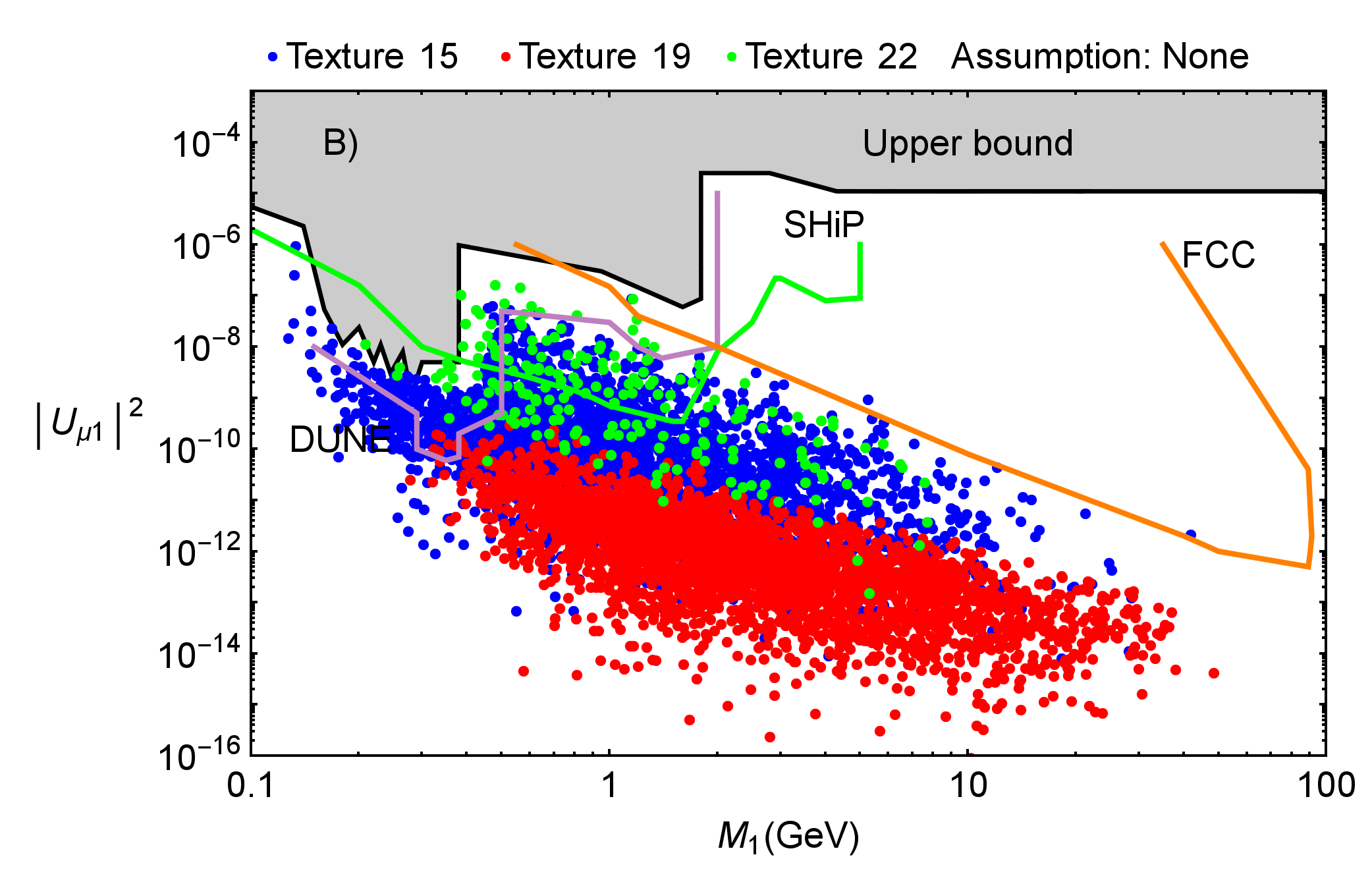}
\end{tabular}
\caption{Similar to \Fig~\ref{fig:individualcasibarand} with the individual mixing elements $|U_{e 1}|^{2}$ (a) and $|U_{\mu 1}|^{2}$ (b) for the lightest sterile neutrino, but showing the predictions for the different texture models.}
\label{fig:individualmixing}
\end{figure*}

In \figu{individualmixing}, the individual mixing elements $|U_{e I}|^{2}$ and $|U_{\mu I}|^{2}$ are shown for the texture sets. It is interesting to observe that the flavor-dependent predictions from the different texture models can be very different. For example, texture~22 produces large mixings in $|U_{\mu1}|^2$, small mixings in $|U_{e1}|^2$, and large total mixings. On the other hand, texture~15 tends to produce larger mixings in both channels. This example demonstrates that the information from different channels can be used as a model discriminator.

\section{Summary and conclusions}

In this work, we have studied the current experimental bounds and future experimental sensitivities to the GeV seesaw models with three sterile neutrinos, which have the advantage that no fine-tuning of the masses is needed for successful leptogenesis. Compared to models with two neutrinos, such models allow for more parameter space freedom as, for instance, the seesaw bound does not apply. Consequently, the parameter space reach of future experiments will be more limited, although we find that models with extremely small mixings seem to require some fine-tuning.

As far as the predictions from theory are concerned, we have first of all studied the Casas-Ibarra and random matrix cases, which can in principle reproduce the same parameter space. However, note that the Casas-Ibarra parametrization uses the observed lepton mixing angles and neutrino masses as an input, which means that it automatically satisfies these constraints. Since especially the parameter space for large mixings seems to require some fine-tuning in the mixing matrix entries, the random case tends to favor smaller (but not extremely small) mixings. \vspace{-0.1cm}

As another example, we have studied the predictions from flavor symmetry models -- with interesting observations. First of all, the flavor symmetry can be used to control if larger or smaller mixings with the sterile neutrinos are produced. Maybe even more interesting, we have shown that different channels sensitive to $|U_{eI}|^2$ and $|U_{\mu I}|^2$ provide complementary information, which can be used as a model discriminator. We therefore encourage the experimental collaborations to study the sensitivities to different channels in order to have independent information on the individual mixings and the total mixing with the sterile neutrinos. We have also demonstrated that different experiments are complementary, in the sense that, for example, FCC can test the heaviest HNL mass in many models in which the lightest HNL mass cannot be accessed in DUNE or SHiP and vice versa. \vspace{-0.1cm}

Regarding  current and future bounds, we have encountered subtleties in their interpretation, and we have highlighted the importance to compare them under equal assumptions. For example, often the sensitivity to the total mixing with the sterile neutrinos $|U_{I}|^2$ is shown, which can, for the leading channels in SHiP and DUNE, only be derived under certain assumptions for the ratios of the flavor-dependent active-sterile mixings. These assumptions have to be applied to both the bounds and sensitivities in the same way  to assess the future parameter space reach. A more appropriate representation for these experiments might be  to show the sensitivity to the individual mixings $|U_{eI}|^2$ and $|U_{\mu I}|^2$, whereas FCC is directly sensitive to $|U_{I}|^2$.

In this study, we only considered normal neutrino mass ordering, but we expect that similar arguments apply to the inverted ordering. We have chosen three generations of  sterile neutrinos at the GeV scale, motivated by symmetry to the active ones and by avoiding fine-tuning of the masses to allow for successful leptogenesis. Note, however, that two generations of sterile neutrinos are sufficient to explain the two mass square differences observed in the neutrino oscillation data, and, in fact, the parameter space will be more strongly constrained. On the other hand, the predictive power of three generation models at the GeV scale has been limited in generic approaches, while flavor symmetries have been shown to reduce this freedom and increase the predictability of the parameters. There is no dark matter candidate in our models; but the models could be extended by adding a fourth weakly coupled neutrino at the keV scale. The mixing with the active neutrinos must be small such that its lifetime is long enough on cosmological scales to match the observed dark matter abundance. 

\subsection*{Acknowledgments}
We appreciate discussions and comments with Heiko Lacker and Heinrich P{\"a}s.

\clearpage
\bibliographystyle{apsrev4-1}
%\bibliography{references}

\end{document}